\documentclass[journal]{IEEEtran}
\usepackage{amsmath,amsfonts}
\usepackage{amssymb}
\usepackage{array}
\usepackage{lipsum}
\usepackage{orcidlink}
\usepackage[caption=false,font=normalsize,labelfont=sf,textfont=sf]{subfig}
\usepackage{subcaption}
\usepackage{algorithm}
\usepackage{breqn}
\usepackage{textcomp}
\usepackage{siunitx}
\usepackage{stfloats}
\usepackage{url}
\usepackage{verbatim}
\usepackage{graphicx}
\usepackage{subcaption}
\usepackage{amsmath}
\usepackage{mathrsfs}
\usepackage{booktabs}
\usepackage{multirow}
\usepackage{cite}
\usepackage{xurl} %
\usepackage{soul}
\usepackage{xcolor}
\sethlcolor{yellow} 
\usepackage{booktabs}
\usepackage{tabularx}
\usepackage{algorithmicx}
\usepackage{caption}   

\begin{document}


\title{iBEAMS: A Unified Framework for Secure and Energy-Efficient \textbf{I}SAC-MIMO Systems leveraging \textbf{B}ayesian \textbf{E}nhanced Learning, and \textbf{A}daptive Game-Theoretic \textbf{M}ulti-Layer \textbf{S}trategies}

\author{Mehzabien Iqbal ~\IEEEmembership{Graduate Student Member,~IEEE}, and Ahmad Y. Javaid ~\IEEEmembership{Senior Member,~IEEE}}

\maketitle

\begin{abstract}
Next generation ISAC networks operating in the mmWave and THz bands must provide physical layer secrecy against potential eavesdroppers (mobile and static) while coordinating distributed hybrid edge nodes under stringent power and QoS constraints. However, these requirements are rarely addressed in a unified manner in existing ISAC physical layer security designs. This paper proposes iBEAMS, a hierarchical Stackelberg–GNE–Bayesian framework for secure and energy efficient ISAC with distributed hybrid nodes. The proposed architecture integrates: (i) a Stackelberg leader at the ISAC base station that jointly optimizes total transmit power, power splitting among confidential data, artificial noise, and sensing, and broadcasts incentive prices to shape follower utilities; (ii) a Generalized Nash Equilibrium Game in which hybrid nodes select transmit powers and transmission versus jamming roles under coupled interference constraints and base station-imposed leakage penalties; and (iii) a Bayesian cooperative refinement layer that forms geometry-aware jamming coalitions aligned with the posterior distribution of the eavesdropper's Angle of Arrival. Simulations over carrier frequencies from 28 GHz to 3 THz demonstrate hierarchical convergence of both base station and hybrid node decisions with stable cooperative friendly jamming. iBEAMS attains approximately 4.4--4.7 bps/Hz average secrecy rate, achieves about  higher Secrecy Energy Efficiency (SEE), and delivers 30--70\% higher SEE than a Stackelberg-decision-based baseline, while maintaining zero outage at 28 GHz. Moreover, the posterior-aligned jamming remains sharply directive and resilient under mobile eavesdroppers and increasing adversary density, indicating that iBEAMS can simultaneously act against static and mobile adversaries while coordinating hybrid edge nodes under limited power and QoS constraints.

\end{abstract}

\section{Introduction}
\label{Intro}
\IEEEPARstart{T}{he} need for secure, energy-efficient, and sensing-aware wireless infrastructure is rapidly intensifying as beyond fifth-generation (B5G) and sixth-generation (6G) networks are envisioned to support safety-critical, autonomous, and immersive applications while still providing high-throughput data pipes \cite{chowdhury20206g, akyildiz20206g}. To achieve this vision, next generation communications increasingly rely on millimeter-wave (mmWave) and terahertz (THz) bands to deliver multi gigabit transmission in dense network deployments \cite{10227884,10608250}. However, the same highly directional beams, large bandwidths, and aggressive spatial reuse that enable these performance gains also intensify security exposure due to high propagation loss, beam leakage, reflections, and the inherently broadcast nature of wireless channels \cite{10731915,8845312}. Consequently, as wireless infrastructures increasingly integrate autonomous and Artificial Intelligence(AI)-driven control, cyber-physical threats are becoming more sophisticated, adaptive, and context-aware, thereby creating a pressing need for frameworks that jointly account for security, performance, and situational awareness at the physical layer. However, existing approaches typically treat communication, sensing, and Physical Layer security (PLS) in isolation or rely on static, far-field models and offline optimization, leaving the problem of designing low-complexity, cross-layer algorithms that can jointly secure and manage mmWave/THz systems under near-field operation, mobility, and imperfect channel knowledge open.

Within this context, Integrated Sensing and Communication (ISAC) has emerged as a key paradigm for future wireless systems. By unifying radar and communication functions, ISAC enhances PLS through the joint exploitation of sensing and communication (S\&C) metrics and information-theoretic limits (detection, estimation, efficiency, and reliability), while improving spectral, energy, and hardware efficiency through spectrum sharing, hardware reuse, and integrated signal processing, in contrast to architectures that treat sensing and communication separately \cite{9737357,9761984}. However, ISAC operation in the mmWave/THz regime induces a tightly coupled communication–sensing decision problem in the presence of strategic eavesdroppers and jammers, where legitimate nodes must act under incomplete and noisy knowledge of adversarial behavior \cite{11172669,9606831}. This challenge is especially critical for B5G/6G applications, including vehicular and V2X networks, UAV swarms, industrial IoT, and mission-critical surveillance or defense systems where communication, sensing, and security are deeply intertwined. Eventually, failures can have immediate physical impacts. In such scenarios, ISAC transceivers must continuously sense the environment, infer adversary behavior, and adapt beams, power, artificial noise, sensing effort, and jamming roles in an energy-efficient manner while satisfying stringent latency and reliability requirements\cite{9737357,liu2022survey,9919739}.

Game theory has emerged as a powerful methodology for modeling and designing PLS strategies in ISAC-enabled MIMO systems by capturing the strategic interactions among legitimate transceivers, eavesdroppers, and jammers and providing a rigorous framework to define their performance and analyze stable operating points of the system \cite{10543060,10945754}. In parallel, Bayesian learning–based methods offer a principled way to update beliefs about unknown or time-varying adversaries from ISAC measurements and adapt transmission policies accordingly, enabling joint optimization of sensing, communication, and security decisions rather than treating them as isolated control variables under simplified, quasi-static models.
\vspace{-0.2cm}

\subsection{Related Work}
\label{Rwork}
ISAC has become a key paradigm for B5G/6G since sharing radio resources for sensing and communication can improve spectral and energy efficiency in dense deployments \cite{ ouyang2023integrated, 11080116}. However, practical operation in power, and interference-limited settings defines the importance of security against strategic adversaries and energy-aware resource allocation \cite{10945754}. In this section, we present a detailed overview of the current state of the art of ISAC frameworks.

Game-theoretic and sequential decision-making approaches are increasingly used to secure ISAC under uncertainty. According to \cite{mamaghani2025securing}, a single ISAC base station and a mobile UAV eavesdropper are modeled as a Stackelberg game, where the UAV learns its trajectory through Deep Reinforcement Learning (DRL), and the base station optimizes beamforming and power to balance secrecy, sensing, and energy efficiency. While effective, this setting remains limited to a single BS–UAV interaction and does not jointly incorporate Bayesian belief updates from ISAC measurements or support multi-node, equilibrium-consistent coordination. By contrast, iBEAMS addresses a broader MIMO-ISAC network with multiple legitimate nodes and adversaries, combining Bayesian sensing-driven learning with Stackelberg and GNE formulations to jointly optimize beamforming, power splitting, artificial noise, sensing effort, and hybrid-node jamming/relaying roles.In addition to this, \cite{wang2025safeguarding} introduces only Bayesian game formulations to handle incomplete information about adversary locations or strategies, ensuring robust transmission policies in uncertain environments. In \cite{10897852}, secure hybrid beamforming for mmWave ISAC is studied using a partially connected, dynamic sub-array architecture, where transmit power is minimized subject to secrecy-rate and sensing beampattern constraints under perfect or imperfect CSI (Channel State Information) through a two-stage design using SDR/SCA and the S-procedure. In contrast, iBEAMS generalizes beyond scenario-specific beamforming by integrating Bayesian sensing-driven belief updates with game-theoretic security modeling to support strategic-adversary awareness and scalable multi-agent, multi-layer PLS coordination with learning-enabled adaptation.
At the system level, researchers have also investigated advanced enablers for secure ISAC, such as reconfigurable intelligent surfaces (RIS), near-field MIMO, and AI-driven optimization frameworks \cite{10890949, 10870062}. Nonetheless, the dependence on static uncertainty sets and approximation-based optimization exposes a key limitation, as such methods lack the adaptive, belief-driven, and dynamically reconfigurable robustness that iBEAMS seeks to provide.

Therefore, prior studies have established the potential of sensing aided beamforming, game theoretic PLS modeling, and system level enablers for secure ISAC. However, most existing designs remain largely static, optimize isolated objectives, and do not jointly capture adversarial uncertainty, cooperative jamming, and entropy aware sensing, ultimately addressing security and energy efficiency together. The key open problem is to co-design base station power splitting, distributed node power and role control, and adaptive sensing to maximize secrecy and energy efficiency against mobile adversaries with partial or unknown CSI under strict power and QoS constraints. This motivates iBEAMS, which unifies Bayesian posterior learning, Stackelberg base station control, GNE based hybrid node interactions, and learning driven policy optimization within a single three layer architecture for secure and energy efficient MIMO ISAC networks.

\subsection{Contributions}
Driven by recent progress in ISAC, this work advances ISAC MIMO systems by addressing key challenges that jointly impact PLS, sensing quality, and energy efficiency. Fig.~\ref{iBEAMS_unified framework} summarizes the overall iBEAMS concept and key components of the proposed intelligent framework. The key contributions are as follows:
\begin{itemize}
\item \textbf{Primary contribution: iBEAMS as a unified hierarchical framework for secured and energy efficient ISAC:}
We propose \emph{iBEAMS}, a unified three-layer Stackelberg–GNE–Bayesian architecture for secure and energy-aware ISAC, where the base station acts as a Stackelberg leader, Hybrid Nodes (HNs) as strategic followers, seeking to reach a Generalized Nash Equilibrium (GNE), and a Bayesian refinement layer forms sensing-driven cooperative jamming coalitions. By iteratively updating beliefs and PLS metrics from ISAC measurements each slot, the framework jointly coordinates power, beamforming, sensing, and node role switching.  To the best of our knowledge, this is the first ISAC–PLS architecture that integrates Stackelberg leadership, multi-agent GNE interaction, and Bayesian inference within a single hierarchical game theoretic model.

\item \textbf{Layer 1: ISAC BS as Stackelberg leader with Bayesian belief feedback:}
In the first layer, the ISAC base station (BS) coordinates and broadcasts prior-channel PLS metrics and the eavesdroppers’ Angle-of-Arrival (AoA) belief as part of a Stackelberg leader strategy. The ISAC BS jointly adapts its total transmit power and power split ratios for confidential data $\alpha$, artificial noise (AN) $\beta$ and sensing $\gamma$ based on secrecy feedback and posterior entropy; sensing and AN are increased when secrecy degrades or uncertainty is high, and reduced when the posterior is confident that it will avoid excess RF consumption. In parallel, incentive prices ($\pi$, $\tau$, and $\zeta$) are broadcast to incorporate leakage and energy penalties into the HN utilities, resulting in an incentive-compatible and entropy-aware power and splitting policy aimed at maximizing SEE. 

\item \textbf{Layer 2: Hybrid-node GNE for distributed power and role selection:}
In the second layer, hybrid nodes operate as strategic followers and solve a Generalized Nash Equilibrium (GNE) under coupled interference and shared power constraints. Given the leader’s broadcast power splits and incentive prices, each node optimizes a local utility that trades off secrecy benefits, transmit power costs, and leakage penalties. The resulting equilibrium yields a distributed power allocation and an initial role assignment (transmitting node - THN, vs. jamming node - JHN), providing an incentive-consistent and energy efficient coordination mechanism.

\item \textbf{Layer 3: Bayesian cooperative refinement for geometry-aware jamming:}
In the third layer, HNs cooperate using the updated eavesdropper AoA posterior and its entropy to refine the jamming configuration. Coalitions are formed toward the most likely eavesdropper directions, prioritizing nodes that are geometrically aligned with these angles as focused jammers, while reallocating power to suppress redundant or low-impact jammers. The refinement iterates until the secrecy metric stabilizes, thus establishing a posterior-aligned, geometry-aware jamming strategy that concentrates energy where it is most effective, preserves legitimate QoS, and improves SEE.
\end{itemize}
\subsection{Organization}
The remainder of this paper is organized as follows. Section~\ref{sys_model} discusses the system model of iBEAMS. Section~\ref{problemform} formulates the core optimization problems underlying ISAC operation, PLS, and energy efficiency. Building on these formulations and associated constraints, section~\ref{game_theory Framework} introduces the proposed iBEAMS architecture, which is described as a three-layer hierarchical game-theoretic control framework. Section~\ref{num_eval} details the simulation setup and provides the corresponding numerical evaluation. Finally, Section~\ref{Conc.} concludes the article with key findings and outlines promising directions for future research.

\section{System Architecture}
\label{sys_model}
\begin{figure}[htp]
\centering
\includegraphics[width=0.48\textwidth]{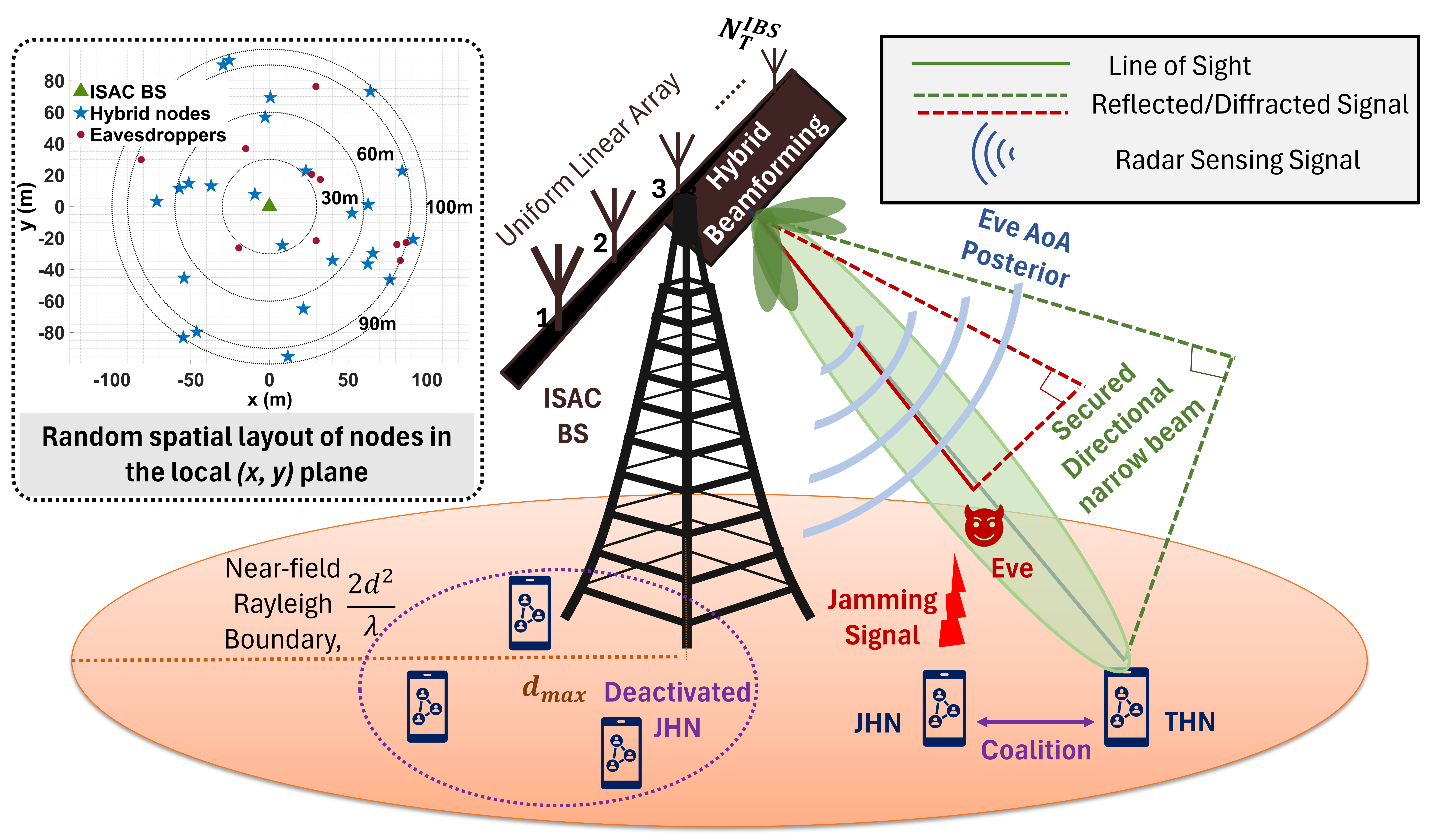}
\caption{System Model of Unified iBEAMS Framework}
\label{sys_modelfig}
\end{figure}
This section provides the rigorous system architecture and is organized as follows: Section~\ref{net_model} presents the network model; Section~\ref{antennaandRF} details the antenna arrays and RF-chain configuration; Section~\ref{signal_model} introduces the propagation channel model and noise model; and Section~\ref{powersplit_model} formulates the energy-aware power allocation and sensing model. In line with that, Figure~\ref{sys_modelfig} illustrates the proposed iBEAMS network architecture, which consists of an ISAC-enabled base station (IBS), a set of distributed hybrid nodes (HNs), and potential adversaries/eavesdroppers (Eves).
\subsection{Network Model}
\label{net_model}
To develop a network model, we consider a single-cell ISAC network operating over a 5G NR style TDD downlink, where a single ISAC base station (IBS) serves multiple hybrid edge nodes (HNs) in the presence of eavesdroppers (Eves). The IBS is located at the center of a circular radius service region $d_{\max}$, with randomly distributed HNs and Eves potentially appearing near the IBS and toward the cell edge (Fig.~\ref{sys_modelfig}). In our network model, link formation and performance evaluation are driven by a 3GPP-compliant propagation and near-field channel construction \cite{zhu20213gpp,poddar2025overview}. 
\vspace{0.2mm}
\subsection{Antenna Arrays, Steering Vectors, and Hybrid Precoding}
\label{antennaandRF}
To support highly directional mmWave/THz transmission under practical RF chain constraints, the ISAC base station employs a large-scale array with $N_T^{\mathrm{IBS}}$ transmit antennas and $N_{\mathrm{RF}}$ RF chains using a partially connected hybrid-precoding architecture. The downlink precoder at slot $t$ is modeled from~\cite{iqbal2026dual}, thus
\begin{equation}
\begin{aligned}
\mathbf{F}(t) &= \mathbf{F}_{\mathrm{RF}}(t)\mathbf{F}_{\mathrm{BB}}(t),\\[-3pt]
&\mathbf{F}_{\mathrm{RF}}(t)\!\in\!\mathbb{C}^{N_T^{\mathrm{IBS}}\!\times\! N_{\mathrm{RF}}},
\ \mathbf{F}_{\mathrm{BB}}(t)\!\in\!\mathbb{C}^{N_{\mathrm{RF}}\!\times\! N_s},
\ N_s\!\le\! N_{\mathrm{RF}}.
\end{aligned}
\label{eq:downlink_precoder}
\end{equation}
where $\mathbf{F}_{\mathrm{RF}}(t)$ is the analog RF precoder implemented through constant-modulus phase shifters, and $\mathbf{F}_{\mathrm{BB}}(t)$ is the low-dimensional digital baseband precoder for stream multiplexing, interference management, and power allocation~\cite{6717211,iqbal2026dual}. Under the partially connected structure, each RF chain drives a dedicated sub-array, so $\mathbf{F}_{\mathrm{RF}}(t)$ is approximately block diagonal, achieving a large array gain with substantially fewer RF chains than antenna elements.

Within iBEAMS, RF precoding is realized through phase-only weights aligned with the instantaneous spatial channels of the scheduled HNs, consistent with practical hybrid beamforming implementations that produce highly directive beams over the desired angular sector~\cite{9583918}. To enhance PLS, AN is projected onto the approximate null-space of the aggregate legitimate channel to degrade eavesdropper reception while minimally affecting intended links~\cite{4543070}.

\noindent\textbf{Antenna Geometry and Array Response:}
To characterize the spatial response, standard 3D array-geometry models are adopted for both the IBS and HNs. Each transceiver is equipped with a uniform linear array (ULA) with half-wavelength spacing, $d=0.5\lambda$, which suppresses grating lobes and enables well-conditioned beam steering. A narrowband plane wave arriving from azimuth-elevation $(\alpha_{\mathrm{az}},\epsilon_{\mathrm{el}})$ is associated with the unit direction vector. According to~\cite{balanis2016antenna}
\begin{equation}
\mathbf{u}(\alpha_{\mathrm{az}},\epsilon_{\mathrm{el}})=
\begin{bmatrix}
\cos\epsilon_{\mathrm{el}}\cos\alpha_{\mathrm{az}}\\
\cos\epsilon_{\mathrm{el}}\sin\alpha_{\mathrm{az}}\\
\sin\epsilon_{\mathrm{el}}
\end{bmatrix}.
\label{eq:umatrix}
\end{equation}
For a ULA of $N$ elements oriented along the $y$-axis, the $n$th element location, ${r}_n$ can be expressed  according to~\cite{mailloux2017phased}
\begin{equation}
\mathbf{r}_n=
\begin{bmatrix}
0\\
nd\\
0
\end{bmatrix},\quad
n\in\left\{-\tfrac{N-1}{2},\ldots,\tfrac{N-1}{2}\right\}.
\label{position_nelement}
\end{equation}
Let $k_0=2\pi/\lambda$ denote the free-space wavenumber. Under the narrowband assumption, the phase progression at element $n$ is
\begin{equation}
\phi_n(\alpha_{\mathrm{az}},\epsilon_{\mathrm{el}})
= k_0\,\mathbf{r}_n^\top\mathbf{u}(\alpha_{\mathrm{az}},\epsilon_{\mathrm{el}})
= \pi n \cos\epsilon_{\mathrm{el}}\sin\alpha_{\mathrm{az}},
\end{equation}
where the last equality follows from $d=0.5\lambda$. This phase progression directly links array geometry to angular selectivity.

\noindent\textbf{Steering Vectors:}
Directional transmission and reception are described through array steering vectors. For a ULA with $N$ elements, the steering vector has been defined from~\cite{iqbal2026dual}
\begin{equation}
\mathbf{a}_{N}(\alpha_{\mathrm{az}},\epsilon_{\mathrm{el}})
= \frac{1}{\sqrt{N}}
\begin{bmatrix}
e^{j k_0 d (-(N-1)/2)\cos\epsilon_{\mathrm{el}}\sin\alpha_{\mathrm{az}}} \\
\vdots \\
e^{j k_0 d n \cos\epsilon_{\mathrm{el}}\sin\alpha_{\mathrm{az}}} \\
\vdots \\
e^{j k_0 d ((N-1)/2)\cos\epsilon_{\mathrm{el}}\sin\alpha_{\mathrm{az}}}
\end{bmatrix}
\end{equation}
Accordingly, the IBS and HN steering vectors are defined as
\begin{equation}
\begin{aligned}
\mathbf{a}_{\mathrm{IBS}}(\alpha_{\mathrm{az}},\epsilon_{\mathrm{el}})
&= \mathbf{a}_{N_T^{\mathrm{IBS}}}(\alpha_{\mathrm{az}},\epsilon_{\mathrm{el}}),\\[-3pt]
\mathbf{a}_{\mathrm{HN,tx}}(\alpha_{\mathrm{az}},\epsilon_{\mathrm{el}})
&= \mathbf{a}_{\mathrm{HN,rx}}(\alpha_{\mathrm{az}},\epsilon_{\mathrm{el}})
= \mathbf{a}_{N_{\mathrm{HN}}}(\alpha_{\mathrm{az}},\epsilon_{\mathrm{el}})
\end{aligned}
\label{eq:array_response}
\end{equation}

In addition, for a transmit beamforming vector $\mathbf{w}_{\mathrm{BS}}\in\mathbb{C}^{N_T^{\mathrm{IBS}}}$, the IBS array response and normalized power gain toward $(\alpha_{\mathrm{az}},\epsilon_{\mathrm{el}})$ are
\begin{equation}
\begin{aligned}
A_{\mathrm{IBS}}(\alpha_{\mathrm{az}},\epsilon_{\mathrm{el}})
&= \mathbf{w}_{\mathrm{BS}}^{\mathrm{H}}
   \mathbf{a}_{\mathrm{IBS}}(\alpha_{\mathrm{az}},\epsilon_{\mathrm{el}}),\\[-3pt]
G_{\mathrm{IBS}}(\alpha_{\mathrm{az}},\epsilon_{\mathrm{el}})
&= \!\left|A_{\mathrm{IBS}}(\alpha_{\mathrm{az}},\epsilon_{\mathrm{el}})\right|^{\!2}.
\end{aligned}
\label{eq:ibs_gain}
\end{equation}
Analogous expressions apply for the HNs using $\mathbf{a}_{\mathrm{HN,tx}}(\cdot)$ and $\mathbf{a}_{\mathrm{HN,rx}}(\cdot)$. Coupled with the hybrid precoder in~\eqref{eq:downlink_precoder}, these steering vectors constitute the fundamental building blocks for beam alignment and geometric channel characterization in mmWave/THz ISAC systems~\cite{10897852,8683591,8052157}.

\subsection{Propagation Model}
\label{signal_model}
In this iBEAMS framework, the link formation and performance evaluation are driven by a 3GPP-compliant propagation and near-field channel construction. Specifically, large-scale propagation is generated using the 3GPP TR~38.901 model \cite{zhu20213gpp,poddar2025overview}, incorporating log-distance path loss and lognormal shadowing, while small-scale fading is modeled as Rician to reflect LOS-dominant mmWave/THz links \cite{hadimani2025performance}.
\subsubsection{Near-Field Rician BS–HN and BS–Eve Channels}
\label{sec:channel_model}
 The proposed iBEAMS architecture is instantiated over a geometry-based, near-field MIMO channel model tailored to mmWave/THz ISAC deployments. According to \ref{antennaandRF}, and \cite{balanis2016antenna, en18112966, Bjrnson2021APO}, the effective aperture,  $D_{\mathrm{ap}}$, can be depicted when $f_c$ is the carrier frequency, $\lambda = c/f_c$ the wavelength, and an IBS equipped with a ULA of $N_T^{\mathrm{IBS}}$ antenna elements and inter-element spacing $d_{\mathrm{ant}}$ as
\begin{equation}
    D_{\mathrm{ap}} = (N_T^{\mathrm{IBS}} - 1)\,d_{\mathrm{ant}},
\end{equation}
which induces a Fraunhofer distance,  $ r_F = \frac{2D_{\mathrm{ap}}^2}{\lambda}$, that can be comparable to the cell radius at high carrier frequencies and large array sizes. As a consequence, ISAC BS–node channels are considered using spherical-wave phases. Therefore, the BS–HN link for the $u$th hybrid node (HN) is modeled as a near-field Rician vector where $K_{\mathrm{R}}$ is the Rician $K$–factor.
\begin{equation} \mathbf{h}_{\mathrm{BS},u} = \sqrt{\frac{K_{\mathrm{R}}}{K_{\mathrm{R}}+1}}, \mathbf{h}_{\mathrm{NLOS}}{u}, \label{eq:rician_hn} \end{equation}
Moreover, the LOS (Line of Sight) component incorporates spherical-wave propagation:
where $d_u$ is an effective link distance, such as the average of $\{d_n(\mathbf{x}_u)\}$) and the NLOS component is modeled as spatially white Rayleigh fading model,
\begin{equation}
\begin{aligned}
\big[\mathbf{h}^{\mathrm{LOS}}_{u}\big]_n
&= g(d_u)\,
   \frac{\exp\!\left(-j\frac{2\pi}{\lambda}d_n(\mathbf{x}_u)\right)}
        {\sqrt{N_T^{\mathrm{IBS}}}},
\quad n = 1,\ldots,N_T^{\mathrm{IBS}},\\[-2pt]
\mathbf{h}^{\mathrm{NLOS}}_{u}
&\sim \mathcal{CN}\!\left(\mathbf{0},\frac{1}{N_T^{\mathrm{BS}}}\mathbf{I}\right)
\end{aligned}
\label{eq:combined_los_nlos}
\end{equation}

In our system, HNs channels are fixed over the considered time horizon, capturing quasi-static large-scale propagation with a representative small-scale propagation. However, the BS–Eve channel to the $e$th eavesdropper in slot $t$ follows the same near-field Rician structure but incorporates time variation in the scattered component:
\begin{equation}
    \mathbf{h}_{\mathrm{BS},e}(t)
    = \sqrt{\frac{K_{\mathrm{R}}}{K_{\mathrm{R}}+1}}\,
      \mathbf{h}^{\mathrm{LOS}}_{e}
      + \sqrt{\frac{1}{K_{\mathrm{R}}+1}}\,
      \mathbf{h}^{\mathrm{NLOS}}_{e}(t),
    \label{eq:rician_eve}
\end{equation}
where $\mathbf{h}^{\mathrm{LOS}}_{e}$ is determined by geometry through Fraunhofer distance, $r_f$, while $\mathbf{h}^{\mathrm{NLOS}}_{e}(t)$ is independently incorporated across slots to emulate mobility and fast fading.

\subsubsection{Large-Scale Path Loss and Shadowing}
Let $\mathbf{r}_n \in \mathbb{R}^3$ denote the position of the $n$th BS antenna element and $\mathbf{x} \in \mathbb{R}^3$ the location of a node that can be an HN or an eavesdropper. The element-wise distances are
\begin{equation}
    d_n(\mathbf{x}) = \|\mathbf{r}_n - \mathbf{x}\|_2, 
    \qquad n = 1,\ldots,N_T^{\mathrm{IBS}}.
    \label{eq:elem_distance}
\end{equation}
In addition, the free-space path-loss given by the Friis reference model is widely used in mmWave and THz channel modeling~\cite{rappaport2015millimeter} and in our iBEAMS framework, where distance-dependent large-scale attenuation is captured by a log-distance model with lognormal shadowing.
\begin{equation}
    \mathrm{PL}^{\mathrm{dB}}(d)
    = \mathrm{PL}^{\mathrm{dB}}_{1\mathrm{m}}
      + 10n_{\mathrm{PL}}\log_{10}\!\left(\frac{d}{1~\mathrm{m}}\right)
      + X_{\sigma},
    \label{eq:PL_logdist}
\end{equation}
where $n_{\mathrm{PL}}$ is the path-loss exponent and $X_{\sigma}\sim\mathcal{N}(0,\sigma_{\mathrm{sh}}^2)$ denotes the lognormal shadowing term, consistent with standard large-scale fading models in cellular systems~\cite{goldsmith2005wireless,rappaport2015millimeter}. The corresponding large-scale linear amplitude gain is then
\begin{equation}
    g(d) = 10^{-\,\mathrm{PL}^{\mathrm{dB}}(d)/20},
    \label{eq:pl_gain}
\end{equation} which directly links the stochastic path-loss model to the complex baseband channel coefficients used in the iBEAMS physical-layer simulations.

\subsubsection{Signal Model and Secrecy Metrics}
\label{signal_model}
Building on Section~\ref{antennaandRF} and (\ref{eq:downlink_precoder}), we adopt a hybrid precoding architecture. At each slot $t$, the set of hybrid nodes (HNs) is partitioned into the transmitting/legitimate nodes $\mathcal{U}(t)$ (THNs) and the jamming nodes $\mathcal{J}(t)$ (JHNs), where $\mathcal{U}(t)\cap\mathcal{J}(t)=\emptyset$ and $\mathcal{U}(t)\cup\mathcal{J}(t)$ span all active HNs. This partition determines the effective multiuser downlink signal model, namely the intended signal, multiuser interference, and artificial-noise/jamming leakage observed at legitimate HNs and at the eavesdroppers, based on which we define the instantaneous SINR and the corresponding secrecy-rate metrics.
\paragraph*{Noise Model}
In our system, we have considered that receiver noise accounts for thermal noise and front-end noise figure. With thermal noise power spectral density $N_{0,\mathrm{dBm/Hz}}$, system bandwidth $BW$, and receiver noise figure $NF$, the total noise power at the receiver front-end is
\begin{equation}
    \sigma_n^2
    = 10^{\frac{N_{0,\mathrm{dBm/Hz}} 
    + 10\log_{10}(BW) + NF - 30}{10}}
    \quad [\mathrm{W}],
    \label{eq:noise_power}
\end{equation}
and the additive noise samples satisfy $\mathbf{n}_u(t)\sim\mathcal{CN}(\mathbf{0},\sigma_n^2\mathbf{I})$.

\paragraph*{Received Signal Representation}
Within each slot, the IBS allocates its transmit power across confidential data, AN, and sensing using fractions $(\alpha_t,\beta_t,\gamma_t)$ from the sensing model~\ref{powersplit_model}. Thus, the IBS transmit signal in slot $t$ is modeled as
\begin{equation}
    \mathbf{x}(t) = \mathbf{F}\mathbf{s}(t) 
                  + \mathbf{F}_{\mathrm{AN}}\mathbf{z}(t),
    \label{eq:tx_signal}
\end{equation}
where $\mathbf{F}_{\mathrm{AN}}$ is designed to approximately lie in the null space of the stacked legitimate channels\cite{4543070}, and $\mathbf{s}(t)$ contains the scheduled data streams and $\mathbf{z}(t)$ is the AN vector. Thereby, limiting AN leakage toward THNs while intentionally degrading eavesdropper reception.

The received baseband signal at HN, $u\in\mathcal{U}(t)$ under the combining vector $\mathbf{w}_u$ is
\begin{equation}
    y_u(t)
    =
    \mathbf{w}_u^{\mathrm{H}}
    \Big(
        \mathbf{h}_{\mathrm{BS},u}^{\mathrm{H}}\mathbf{x}(t)
        + \sum_{v\in\mathcal{J}(t)} 
            \tilde{h}_{u\leftarrow v}\,\sqrt{P_v(t)}\,z_v(t)
        + \tilde{n}_u(t)
    \Big),
    \label{eq:rx_signal_hn}
\end{equation}
where $\tilde{h}_{u\leftarrow v}$ denotes an effective channel gain from jammer $v$ to HN $u$, $P_v(t)$ is the jamming power of node $v$, and $\tilde{n}_u(t)$ is the effective noise after combining. The corresponding Eve observation for decoding user $u$ is defined analogously using $\mathbf{h}_{\mathrm{BS},e}(t)$ and an eavesdropper combiner $\mathbf{w}_e$.

\paragraph*{Instantaneous SINR and Secrecy Rate}
Let $\mathbf{f}_u$ denote the effective IBS beam associated with user $u$ (the $u$th column of $\mathbf{F}$), and let $P_u(t)$ denote the power allocated to that stream (discussed in \ref{powersplit_model}). Therefore, the instantaneous SINR at legitimate HN $u$ is
\begin{equation}\small
\mathrm{SINR}_{\mathrm{hn},u}(t)=
\frac{P_u(t)\,|g_{hn,u}(t)|^2}
{\sum_{k\neq u} P_u(t)\,|g_{hn,u}(t)|^2
+ I_u^{\mathrm{J}}(t)+I_u^{\mathrm{AN}}(t)+\sigma_n^2},
\label{eq:SINR_hn}
\end{equation}
where $g_{hn,u}(t)\triangleq \mathbf{w}_u^{H}\mathbf{h}_{\mathrm{BS},u}^{H}(t)\mathbf{f}_k(hn)$, and  $I_{u}^{(\mathrm{J})}(t)$ captures the aggregate interference generated by cooperative jammers and $I_{u}^{(\mathrm{AN})}(t)$ accounts for residual AN leakage toward user $u$. Similarly, for an eavesdropper $e$, the SINR associated with decoding the stream intended for user $u$ is
\begin{equation}\small
\mathrm{SINR}_{e,u}(t)=
\frac{P_u(t)\,|g_{e,u}(t)|^2}
{\sum_{k\neq u} P_u(t)\,|g_{e,u}(t)|^2 + I^{\mathrm{J}}_e(t)+I^{\mathrm{AN}}_e(t)+\sigma_n^2}
\label{eq:SINR_e}
\end{equation}, where $g_{e,u}(t)\triangleq \mathbf{w}_e^{H}\mathbf{h}^{H}_{\mathrm{BS},e}(t)\mathbf{f}_u(t)$, and $I_{e}^{(\mathrm{J})}(t)$ and $I_{e}^{(\mathrm{AN})}(t)$ denote, respectively, the contributions of cooperative-jamming and AN in Eve $e$.

In accordance with the proposed system architecture, we adopt a worst-case leakage model by selecting the most detrimental eavesdropper in each slot:
\begin{equation}
    \mathrm{SINR}_{\mathrm{e},u}^{\max}(t)
    = \max_{e\in\mathcal{E}}\,
    \mathrm{SINR}_{\mathrm{e},u}^{(e)}(t),
    \label{eq:worst_eve}
\end{equation}
The instantaneous secrecy rate then follows the standard Gaussian wiretap formulation~\cite{4557589,wyner1975wire}:
\begin{equation}
    R_{s,u}(t)
    = \Big[\log_2\!\big(1+\mathrm{SINR}_{\mathrm{hn},u}(t)\big)
        -\log_2\!\big(1+\mathrm{SINR}_{\mathrm{e},u}^{\max}(t)\big)
    \Big]^+
    \label{eq:secrecy_rate}
\end{equation}, reflecting the fundamental physical-layer performance metric that is subsequently aggregated and fed back into the multi-layer iBEAMS control framework for power allocation, role selection, and sensing adaptation.

\subsection{Energy- Efficient Power Allocation and Sensing Model}
\label{powersplit_model}
In the proposed iBEAMS architecture, energy awareness is considered through a per–time-slot power-consumption model that accounts for both radiated transmit power and hardware-dependent overhead. Let $P_{\mathrm{BS}}(t)$ denote the instantaneous transmit power of the ISAC BS in slot $t$, and let $P_{u}(t)$ denote the transmit power of HN. Therefore, the aggregate radiated power in slot $t$ is
\begin{equation}
    P_{\mathrm{tx,tot}}(t)
    = P_{\mathrm{BS}}(t) + \sum_{u=1}^{U_{\mathrm{HN}}} P_{u}(t),
    \label{eq:ptx_tot}
\end{equation}
Also, the corresponding total power consumption in slot $t$ is modeled as
\begin{equation}
    P_{\mathrm{slot}}(t)
    = P_{\mathrm{cons}} + \frac{P_{\mathrm{tx,tot}}(t)}{\eta_{\mathrm{pa}}},
    \label{eq:pslot}
\end{equation}
where $P_{\mathrm{cons}} = N_{\mathrm{RF}} P_{\mathrm{rf}} + P_{\mathrm{bb}}$ accounts for the static power drawn by RF chains $P_{\mathrm{rf}}$ and baseband processing $P_{\mathrm{bb}}$ and $\eta_{\mathrm{pa}} \in [0,1]$ denotes the effective power-amplifier efficiency. This affine model is consistent with standard Secrecy Energy Efficiency (SEE) formulations in secure MIMO systems\cite{8709756,8743402}. Thus, SEE in slot $t$ is defined as the ratio between the aggregate secrecy throughput and the total power.
\begin{equation}
    \mathrm{SEE}(t)
    = \frac{\displaystyle \sum_{u \in \mathcal{U}(t)} R_{s,u}(t)}
           {P_{\mathrm{slot}}(t)}
    \quad \text{[bits/Joule]},
    \label{eq:see_def}
\end{equation}
Where, $R_{s,u}(t)$ is the instantaneous secrecy rate as given by~\eqref{eq:secrecy_rate}, and $\mathcal{U}(t)$ denotes the set of HNs scheduled as legitimate receivers in slot $t$. The metric in~\eqref{eq:see_def} quantifies the fundamental trade-off between the enhancement of secrecy and the expenditure of energy.
In accordance with the ISAC signal model in Section~\ref{signal_model}, the BS transmit power in slot $t$ is partitioned among confidential communication, AN, and sensing as
\begin{equation}
    P_{\mathrm{BS}}(t)
    = P_{\mathrm{BS}}^{\mathrm{comm}}(t)
    + P_{\mathrm{BS}}^{\mathrm{AN}}(t)
    + P_{\mathrm{BS}}^{\mathrm{sens}}(t),
    \label{eq:pbs_split}
\end{equation}
with
\begin{equation}
\small
\begin{aligned}
P_{\mathrm{BS}}^{\mathrm{comm}}(t)=\alpha_t P_{\mathrm{BS}}(t),\quad
P_{\mathrm{BS}}^{\mathrm{AN}}(t)=\beta_t P_{\mathrm{BS}}(t),\quad
P_{\mathrm{BS}}^{\mathrm{sens}}(t)=\gamma_t P_{\mathrm{BS}}(t)
\end{aligned}
\label{eq:alphabetagamma}
\end{equation}
where $(\alpha_t,\beta_t,\gamma_t)$ are nonnegative power-splitting coefficients satisfying
\begin{equation}
    \alpha_t + \beta_t + \gamma_t = 1, \qquad \forall t.
    \label{eq:alphasum}
\end{equation}
Here, $\alpha_t$ controls the fraction of BS power devoted to secure communications, $\beta_t$ specifies the AN power injected to mitigate eavesdroppers, and $\gamma_t$ regulates the power reserved for sensing beams used for AoA refinement and environment probing in line with ISAC operation. 
To explicitly model the sensing functionality, a power-normalized sensing beam $\mathbf{w}_{\mathrm{S}}\in\mathbb{C}^{N_T^{\mathrm{IBS}}}$ is steered toward a candidate direction $(\phi_{\mathrm{IBS}}, \theta_{\mathrm{IBS}})$, and the corresponding sensing response is characterized by
\begin{equation}
    R_{\mathrm{S}}(\phi_{\mathrm{IBS}}, \theta_{\mathrm{IBS}}) = 
    \left| \mathbf{w}_{\mathrm{S}}^{\mathrm{H}} \mathbf{a}_{\mathrm{IBS}}(\phi_{\mathrm{IBS}}, \theta_{\mathrm{IBS}}) \right|^2,
    \label{eq:sensing_response}
\end{equation}
where $\mathbf{a}_{\mathrm{IBS}}(\cdot)$ denotes the IBS array steering vector. The sensing beam can be further shaped through a Hamming-tapered amplitude distribution to control the width of the main-lobe and the level of the side-lobe \cite{iqbal2026dual}, 
\begin{equation}
    \mathbf{w}_{\mathrm{S}}(\tau, \phi) =
    \bigl[(1-\tau)\{N_{T}^{\mathrm{IBS}} + \tau \,\mathbf{h}_{\mathrm{Hamming}}\bigr]
    \odot \mathbf{a}_{\mathrm{IBS}}(\phi),
    \label{eq:hamming_taper}
\end{equation}
where $\tau\in[0,1]$ is a taper-control factor, $\mathbf{h}_{\mathrm{Hamming}}$ denotes a normalized Hamming window across the array elements, and $\odot$ represents the hadamard product~\cite{balanis2016antenna,doly2025adaptive,10841413}. Uniform weighting ($\tau=0$) yields a wider beam with higher side lobes, whereas $\tau=1$ produces a narrower, high-gain beam with enhanced angular resolution. Equations~\eqref{eq:ptx_tot}, \eqref{eq:hamming_taper} thus provide a unified theoretical description of energy-aware power allocation and sensing in iBEAMS. The numerical evaluation in Section~\ref{num_eval} directly instantiates these expressions to quantify SEE and sensing performance under realistic channel, and interference conditions.

\section{Problem Formulation}
\label{problemform}
Conventional ISAC designs often optimize communication or sensing in isolation under ideal CSI (Channel Sate Information) and static beam patterns. In contrast, secure ISAC requires joint power-limited co-design of confidential transmission and sensing, inducing fundamental tradeoffs among reliability, secrecy, sensing accuracy, and energy efficiency. This section formulates the underlying optimization problems and shows how iBEAMS addresses the tri-coupled optimization problems of \emph{secrecy}, \emph{sensing}, and \emph{energy efficiency}.
\subsection{ISAC Power-Coupled Decision Space and Constraints}
 The BS power-splitting model in Section~\ref{powersplit_model} and \eqref{eq:pbs_split}–\eqref{eq:alphasum}, the fractions ${\alpha_t,\beta_t,\gamma_t}$ determine how $P_{\mathrm{BS}}(t)$ is fractionated among secure data transmission, AN, and sensing. On the follower side, $P_k(t)$ denotes the transmit power of HN $u$ in slot $t$, and $\mathcal{U}(t)$ is the set of scheduled legitimate receivers (THNs). Thus, ${P_{\mathrm{BS}}(t),\alpha_t,\beta_t,\gamma_t,{P_u(t)}*u}$ constitutes the instantaneous power-allocation decision vector, from which the total radiated power $P*{\mathrm{tx,tot}}(t)$ and the consumed power in slots $P_{\mathrm{slot}}(t)$ are obtained through \eqref{eq:ptx_tot}–\eqref{eq:pslot}, with $P_{\mathrm{cons}}$ and $\eta_{\mathrm{pa}}$ capturing the circuit overhead, and PA efficiency, respectively. 
 However, these decisions are restricted by a coupled set of constraints.  First, the power-simplex and feasibility constraints in~\eqref{eq:alphasum}, \eqref{eq:alphabetagamma} enforce non-negativity and normalization of $(\alpha_t,\beta_t,\gamma_t)$, as well as minimum sensing and AN levels, while per-HN and network-wide power limits bound $P_u(t)$ and the friendly jamming (FJ) budget. Second, for the secrecy and sensing-entropy constraints, each active HN must achieve a secrecy rate above $R_{\mathrm{th}}$ with bounded leakage, and the posterior entropy $H(\mathbf{p})$ must remain below $H^\ast$.  Finally, CSI uncertainty is supposed to be captured through the bounded-error model, which defines an admissible channel set for robust operation.
In this way, the \emph{decision space} for iBEAMS is intrinsically coupled with power constraints; any change in the BS power or the split $(\alpha_t,\beta_t,\gamma_t)$ simultaneously impacts the feasibility of HN transmit powers, FJ activation, secrecy guaranties, and sensing reliability, providing the basis for the optimization developed in the next section \ref{game_theory Framework}.
\vspace{-0.3cm}
\subsection{Leader-Level Power-Allocation and Pricing Problem}
 The leader’s problem in each slot can be written as:
\begin{equation}\small
\begin{aligned}
\max_{P_{\mathrm{BS}},\,\boldsymbol{\theta}_t}\;&
\mathrm{SEE}(t)-\lambda_{\mathrm{sec}}\!\big(R_s^{\star}-\overline{R}_s(t)\big)_+
-\lambda_H\!\big(H(t)-H^{\ast}\big)_+\\
\text{s.t.}\;& 0\le P_{\mathrm{BS}}(t)\le P_{\mathrm{BS}}^{\max},\;
(\alpha_t,\beta_t,\gamma_t)\ \text{satisfy}\ \eqref{eq:alphasum},\\
& \pi_{\min}\le\pi_t\le\pi_{\max},\;
\tau_{\min}\le\tau_t\le\tau_{\max},\;
\kappa_{\min}\le\kappa_t\le\kappa_{\max},
\end{aligned}
\label{eq:leader_obj}
\end{equation}
where $\overline{R}_s(t)$ is the average secrecy rate over served THNs in slot $t$, $R_s^{\star}$ is a target secrecy level, $H(t)$ is the current sensing-entropy of the AoA posterior, and $H^{\ast}$ is the allowable uncertainty bound. The penalty terms capture secrecy violations and excessive sensing uncertainty.
In our framework, this problem is solved through an  adaptive update of $(\alpha_t,\beta_t,\gamma_t)$ and $(\pi_t,\tau_t,\kappa_t)$ using the previous-slot secrecy, leakage, and entropy KPIs, thereby realizing a Stackelberg leader control as detailed in section~\ref{game_theory Framework} and Algorithm~\ref{Algorithm_1}.
\subsection{Follower-Level HN GNE Power and Role Problem}
Given the leader decisions $(P_{\mathrm{BS}}(t),\alpha_t,\beta_t,\gamma_t,\pi_t,\tau_t,\kappa_t)$, each HN, $u$ acts as a strategic follower that selects its transmit power $P_U(t)$ and role (transmitting node vs.\ jamming node) subject to local power constraints and shared interference coupling. Let $\mathcal{A}$ denote the set of active HNs, and $\mathcal{E}$ the set of potential eavesdroppers. Using the SINR expressions defined in Section~\ref{signal_model}, the secrecy rate for HN $u\in\mathcal{A}$ is $R_s(u)$ from \eqref{eq:Rs_slot_t_detailed}, and for each HN, a representative utility, $U_u$ derived from Appendix \ref{app:follower_utility_gne}

The follower-layer problem is thus to avail the best GNE in order to
\begin{equation}
\begin{aligned}
\text{Find} \{P_u^\star(t)\}_u \in \mathcal{P}, \
U_u\!\left(P_u^\star(t), P_{-u}^\star(t)\right)
&\ge
U\!\left(P_u(t), P_{-u}^\star(t)\right),\\
&\forall\, P_u(t)\in\mathcal{P}_u,\ \forall\, u.
\end{aligned}
\end{equation}
where $\mathcal{P}_u$ encodes local power constraints and $\mathcal{P}$ captures shared interference and energy budgets. In our system, we have addressed this constraints through approximating the GNE through iterative best-response over a discrete power grid, addressed in Section\ref{game_theory Framework}, and Algorithm \ref{Algorithm_2}.
\vspace{-0.2cm}
\subsection{PLS and Secrecy-Outage Metrics Problem}
System-level security in conventional ISAC designs is often characterized using either the average secrecy rate or the worst-case secrecy performance, typically under fixed power splits and static jamming strategies. Such models rarely capture how secrecy degrades probabilistically across users when eavesdroppers move, CSI is imperfect, and the sensing uncertainty persists. To explicitly quantify this vulnerability, we evaluate security in iBEAMS through per-slot secrecy-outage metrics over the active legitimate set
$\mathcal{A}$. Using~\eqref{eq:secrecy_rate}, we define
\begin{align}\small
R_{\min} &= \min_{u\in\mathcal{A}} R_s(u), \\
\overline{R} &= \frac{1}{|\mathcal{A}|}\sum_{u\in\mathcal{A}} R_s(u), \\
P_{\mathrm{out}} &= \frac{1}{|\mathcal{A}|}\sum_{u\in\mathcal{A}} 
\mathbf{1}\{R_s(u) < R_{\mathrm{th}}\},
\label{idealRP}
\end{align}
where $R_{\mathrm{th}}$ is the target secrecy rate, $R_{\min}$ captures the worst served legitimate user,
$\overline{R}$ reflects the average secrecy level over all active HNs, and $P_{\mathrm{out}}$ outage probability quantifies the fraction of users that fail to meet the secrecy target in a given slot. The proposed iBEAMS framework explicitly seeks to control this secrecy-outage behavior so that $P_{\mathrm{out}}$ remains below a prescribed reliability level while jointly maximizing SEE and enforcing the entropy-based sensing constraint. 

\subsection{Sensing-Entropy and Belief-Coupled Sensing Problem}
\label{sensing entropy_problem}
Most existing models neither maintain an explicit belief over eavesdropper directions nor impose principled bounds on tolerable angular uncertainty before triggering sensing or beam adaptation. Consequently, secrecy decisions are often weakly quantified. Therefore, the IBS explicitly needs to maintain a posterior belief over the angular domain of potential eavesdroppers. Let $\Phi = \{\phi_1,\dots,\phi_{N_\phi}\}$ denote the
discrete angle grid and $\mathbf{p} = [p_1,\dots,p_{N_\phi}]^{\top}$ the
corresponding posterior with
\begin{equation}
\sum_{i=1}^{N_\phi} p_i = 1, 
\qquad p_i \ge 0,\; \forall i,
\label{eq:prob_constraint}
\end{equation}
and quantify the residual directional ambiguity using Shannon entropy
\begin{equation}
H(\mathbf{p}) = -\sum_{i=1}^{N_\phi} p_i \log_2 p_i.
\label{eq:entropy_def}
\end{equation}
So, a large $H(\mathbf{p})$ indicates a diffuse belief (high uncertainty in Eve’s direction), whereas a small $H(\mathbf{p})$ reflects a concentrated belief and, thus, improved localization confidence.To couple sensing explicitly with secrecy and power control, iBEAMS constrains the posterior uncertainty through
\begin{equation}
H(\mathbf{p(t)}) \le H^\ast,
\label{eq:entropy_constraint}
\end{equation}
where $H^\ast$ is a design parameter determined by sensing resolution and an acceptable level of situational-awareness. Since this constraint may be violated in noisy or highly dynamic environments, it should be embedded into the optimization through a penalty term, thus,
\begin{equation}
\mathcal{L}_H = \lambda_H \big[H(\mathbf{p}) - H^\ast\big]_+,
\label{eq:entropy_penalty}
\end{equation}
with $\lambda_H > 0$ and $[x]_+ = \max(x,0)$.

Given sensing beamforming weights, $\mathbf{w}_{\mathrm{S}}$ from \eqref{eq:hamming_taper} used to illuminate different angular sectors, the belief-coupled sensing subproblem is then formulated following \eqref{eq:prob_constraint}, and \eqref{eq:entropy_constraint}
\begin{equation}
\min_{\mathbf{p},\,\mathbf{w}_{\mathrm{S}}} \quad
P_{\mathrm{out}}^{\mathrm{sys}} + \lambda_H \big[H(\mathbf{p}) - H^\ast\big]_+,
\label{eq:isac_obj}
\end{equation}
thereby, enforcing a joint design in which secrecy-outage reduction and bounded sensing uncertainty are optimized simultaneously. In the iBEAMS implementation, those problems are addressed through Algorithm \ref{Algorithm_1} and Algorithm \ref{algorithm_3} described in more detail in section~ \ref{game_theory Framework}.

\subsection{Friendly-Jamming (FJ) and Geometry-Aware Refinement}
FJ schemes in secure wireless networks typically rely on pre-selected or static jammer sets, often ignoring (i) how jammer activation interacts with user secrecy rates under dynamic CSI and mobility, (ii) the geometric relationship between legitimate users and eavesdroppers, and (iii) the explicit energy cost of sustaining multi-node jamming. As a result, jamming resources are frequently over-provisioned as wasting power or poorly targeted with insufficient secrecy gain, and the resulting designs do not explicitly optimize SEE.  Let
$\mathsf{Jam}(u)\in\{0,1\}$ denote the binary role of HN $u$, and $n_J = \sum_{u\in\mathcal{A}}\mathsf{Jam}(u)$, the number of active jammers.Thus, a canonical FJ design problem can be written as
\begin{align}\small
\min_{\{\mathsf{Jam}(u)\},\, \mathbf{g}} \quad 
& P_{\mathrm{out}}^{\mathrm{sys}}(\mathsf{Jam}, \mathbf{g}),
\label{eq:FJ_obj}\\
\text{s.t.}\quad 
& \mathsf{Jam}(u) \in \{0,1\}, \ \forall u \in \mathcal{A}, 
\label{eq:FJ_binary}\\
& P_{\mathrm{FJ}} = n_J P_{\mathrm{HN}}^{\max} \leq P_{\mathrm{FJ}}^{\max}, 
\label{eq:FJ_power}\\
& P_{\mathrm{leak}}^{(u)} \le \rho_{\mathrm{leak}} P_{\mathrm{lu}}^{\max}, 
\ \forall u \in \mathcal{A}.
\label{eq:FJ_leak}
\end{align}
where $\mathbf{g}$ is the aggregate jamming beamforming vector, $P_{\mathrm{FJ}}$ is the total jamming power, and $P_{\mathrm{leak}}^{(u)}$ bounds interference leakage toward legitimate users.
Within iBEAMS, this theoretical FJ problem is addressed in two tightly coupled stages, aligned with Algorithms \ref{Algorithm_2} and \ref{algorithm_3}.

\begin{figure*}[htp]
    \centering
\includegraphics[width=0.98\textwidth]{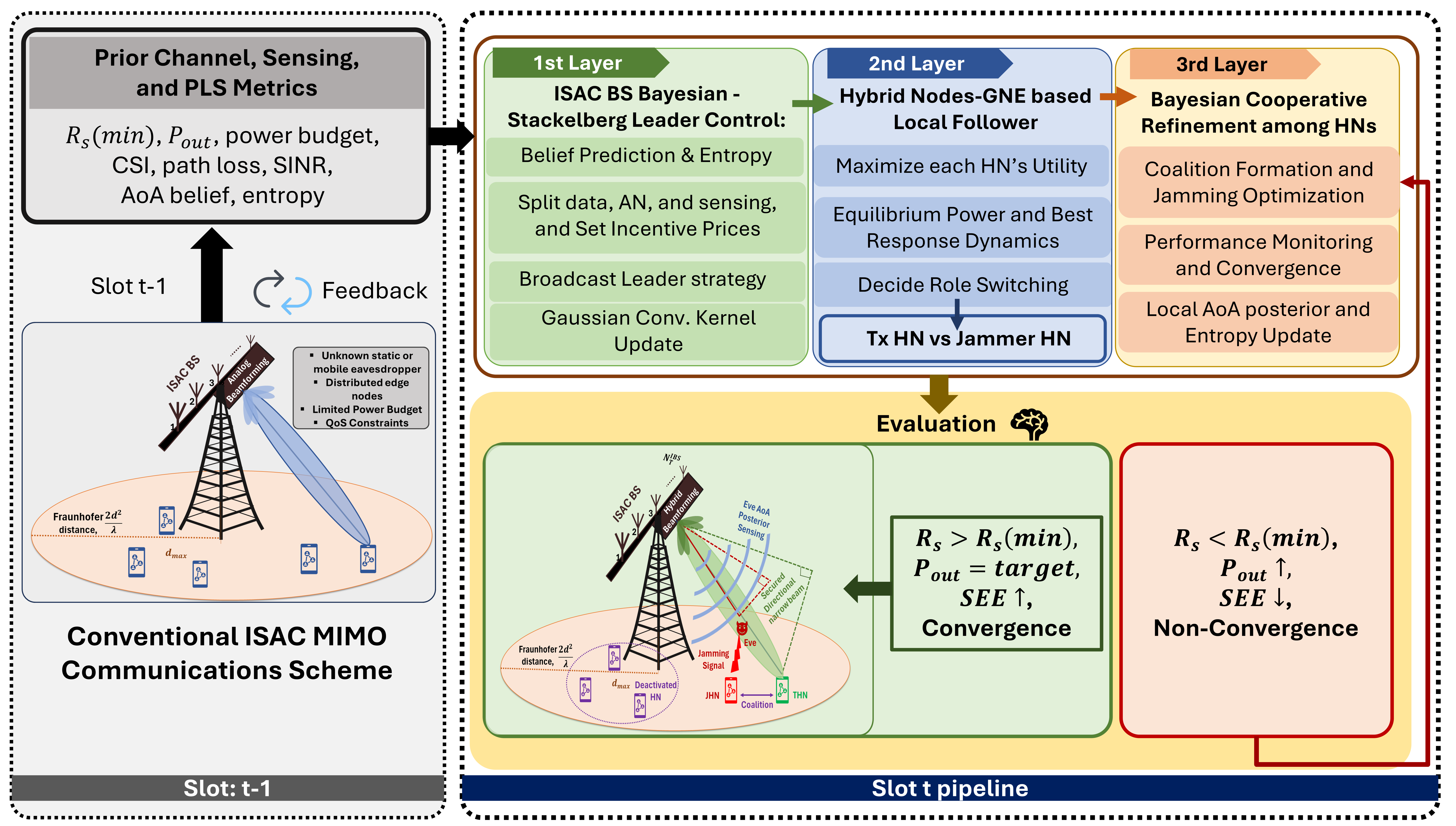}
    \caption{Illustration of Unified Framework of iBEAMS leveraging Bayesian Learning, and Adaptive Game-Theoretic Multi-Layer Strategies of ISAC MIMO Systems}
    \label{iBEAMS_unified framework}
\end{figure*}
\section{iBEAMS Unified Game-Theoretic Framework}
\label{game_theory Framework}
Building on the coupled decision space and constraints in Section~\ref{problemform}, the proposed iBEAMS architecture is modeled as a three-layer hierarchical game-theoretic control framework. In each scheduling slot $t$, the IBS acts as a \emph{Stackelberg-Bayesian leader} (Algorithm~\ref{Algorithm_1}), HNs play a \emph{GNE} game as strategic followers (Algorithm~\ref{Algorithm_2}), and a \emph{Bayesian sensing and refinement layer} updates the AoA belief over potential eavesdroppers and adjusts cooperative jamming coalitions (Algorithm~\ref{algorithm_3}). The game-theoretic principles adopted in this work are inspired by prior contributions in wireless networks and PLS~\cite{9535143,10608156,7903731}. Nevertheless, the central idea and overall framework proposed here are original, as they integrate and extend these concepts into a unified formulation rather than treating each game-theoretic model independently. This three-layer structure directly implements the three-coupled design objectives of \emph{secrecy}, \emph{sensing}, and \emph{energy efficiency} by addressing the power-simplex, secrecy-outage, sensing-entropy, and CSI-uncertainty constraints mentioned in Section~\ref{problemform}. Fig.~\ref{iBEAMS_unified framework} summarizes the concept of iBEAMS and the key components of the proposed intelligent framework.
\begin{algorithm}[t]
\caption{ISAC BS Bayesian-Stackelberg Leader Control}
\label{Algorithm_1}
\textbf{Require:} Prior AoA posterior $p_{t-1}(\theta_E)$ of Eve's AoA; kernel bandwidth $\sigma_{t}$; secrecy KPIs: $\{R_s(t-1), P_{\text{out}}(t-1)\}$; BS power budget $P_{\text{BS}}$;
    incentive bounds $(\pi_{\min},\pi_{\max}),(\tau_{\min},\tau_{\max}),(\kappa_{\min},\kappa_{\max})$;
    secrecy target $R_s^{\text{target}}$; entropy threshold $H^{\max}$.
 \textbf{Ensure:} Adaptive power split $(\alpha_t,\beta_t,\gamma_t)$;
incentive prices $(\pi_t,\tau_t,\kappa_t)$;
predicted belief $p^{-}_{t}(\theta_E)$; updated kernel $\sigma_{t+1}$.

\textbf{Step $1$: Bayesian Belief Prediction}\\
Predict the prior of Eve’s AoA using a Gaussian convolution kernel from \eqref{eq:belief_prediction}
and compute its entropy following \eqref{eq:entropy_predicted}

\textbf{Step $2$: Entropy-Aware Sensing Allocation}\\
Determine sensing portion based on uncertainty:\[
\gamma_t = \Gamma(H_t),\quad 
\Gamma'(\cdot)>0,\; \Gamma(0)=\gamma_{\min},\; \Gamma(H^{\max})=\gamma_{\max}.
\]
\textbf{Step $3$: Secrecy-Deficit-Driven AN Allocation}\\
Compute secrecy error relative to the target:\[e_t = R_s^{\text{target}} - R_s(t-1).\]
Update AN power:
\[\beta_t = \text{limit}(\beta_{t-1} + k_s e_t,\, 0,\,1-\gamma_t).\]

\textbf{Step $4$: Communication Power Adjustment}\\
Set the data power fraction:
\[\alpha_t = 1 - \beta_t - \gamma_t.\]
\textbf{Step $5$: Incentive Price Design (Stackelberg Leader Action)}\\
Based on observed performance:
\begin{align*}
\pi_t   &= \text{limit}\big(\pi_{t-1} + k_\pi ( I_{\text{jam}}^{\text{benefit}} ),\, \pi_{\min},\pi_{\max}\big),\\
\tau_t  &= \text{limit}\big(\tau_{t-1} + k_\tau ( \Xi^{\text{avg}} - \Xi^{\text{target}} ),\, \tau_{\min},\tau_{\max} \big),\\
\kappa_t&= \text{limit}\big(\kappa_{t-1} + k_\kappa (H_t - H^{\max}),\,\kappa_{\min},\kappa_{\max}\big),
\end{align*}
where, $\pi_t$ rewards jamming contributions, $\tau_t$ penalizes leakage, and $\kappa_t$ rewards sensing and information gain.

\textbf{Step $6$: Send Leader Strategy to Nodes}\\
Broadcast $(\alpha_t,\beta_t,\gamma_t)$ and $(\pi_t,\tau_t,\kappa_t)$ to all HNs.

\textbf{Step $7$: Kernel Update}\\
Adapt kernel width for next iteration:
\[\sigma_{t+1} = \sigma_t + \eta_\sigma(H_t - H^{\max}).
\]
Return, $(\alpha_t,\beta_t,\gamma_t,\pi_t,\tau_t,\kappa_t)$ and $p_t^{-}$, $\sigma_{t+1}$.
\end{algorithm}
\subsection{Hierarchical Game Structure and Joint Decision Variables}
\label{gamestructure_decisionvariables}
This subsection formalizes the iBEAMS control architecture as a hierarchical game and specifies the key decision variables at each layer. 
In slot $t$, the leader (IBS) selects the control vector,
\begin{equation}
  \mathbf{a}_t
  =
  \big(
  P_{\mathrm{BS}}(t),\,
  \alpha_t,\beta_t,\gamma_t,\,
  \pi_t,\tau_t,\kappa_t
  \big),
  \label{eq:leader_action_vector}
\end{equation}
where $(\pi_t,\tau_t,\kappa_t)$ are incentive prices broadcast to HNs that enter their utilities as jamming rewards, leakage penalties, and sensing/entropy rewards, respectively.

On the follower side, each HN, $u$ selects a transmit power $P_u(t)$ and a node $\text{role}_u(t)\in\{\text{THN},\text{JHN}\}$, yielding the follower action profile
\begin{equation}
  \mathbf{p}_t
  =\big\{ P_u(t),\,\text{role}_u(t) \big\}_{u\in\mathcal{A}(t)},
  \label{eq:follower_strategy_profile}
\end{equation}
where $\mathcal{A}(t)$ is the set of active HNs in the slot $t$ and $\mathcal{J}(t)\subseteq\mathcal{A}(t)$ is the set of friendly jammers induced. The Bayesian sensing layer maintains a posterior belief over the discrete
angular grid as $\Phi=\{\phi_1,\dots,\phi_{N_\phi}\}$:
\begin{equation}
  \mathbf{p}(t)
  =\big[ p_1(t),\dots,p_{N_\phi}(t) \big]^{\top},
  \qquad
  \sum_{i=1}^{N_\phi} p_i(t) = 1,\quad p_i(t)\ge 0,
  \label{eq:belief_vector_slot_t}
\end{equation}
with entropy
\begin{equation}
  H\big(\mathbf{p}(t)\big)
  = -\sum_{i=1}^{N_\phi} p_i(t)\log_2 p_i(t),
  \label{eq:entropy_slot_t}
\end{equation}
constrained through $H(\mathbf{p}(t))\le H^\ast$ as in
\eqref{eq:entropy_constraint}.

The aggregate radiated power and the total consumed power in slot $t$ follow directly from the power model in section~\ref{powersplit_model}, from \eqref{eq:ptx_tot} and \eqref{eq:pslot}, while the $\mathrm{SEE}(t)$ is defined in \eqref{eq:see_def}. Together, these relations link the leader’s decision variables, $P_{\mathrm{BS}}(t)$ and $\{P_u(t)\}_{u\in\mathcal{U}(t)}$ to the instantaneous energy cost and secrecy performance: any choice of the power-splitting fractions $(\alpha_t,\beta_t,\gamma_t)$, and BS power $P_{\mathrm{BS}}(t)$ determines $P_{\mathrm{tx,tot}}(t)$ and $P_{\mathrm{slot}}(t)$ through \eqref{eq:ptx_tot}–\eqref{eq:pslot}, and hence the SEE term in \eqref{eq:see_def} that appears in the leader’s objective.

In Stackelberg form, the IBS (leader) selects an action vector
$\mathbf{a}_t \in \mathcal{A}_{\mathrm{lead}}$ (including $P_{\mathrm{BS}}(t)$,
$(\alpha_t,\beta_t,\gamma_t)$, and prices $(\pi_t,\tau_t,\kappa_t)$),
anticipating the GNE response
of followers $\mathbf{p}_t^\star(\mathbf{a}_t)$. The resulting hierarchical optimization in slot $t$ can be written as
\begin{equation}
  \max_{\mathbf{a}_t \in \mathcal{A}_{\mathrm{lead}}}
  F\big(\mathbf{a}_t,\mathbf{p}_t^\star(\mathbf{a}_t),\mathbf{p}(t)\big),
  \label{eq:stackelberg_overall}
\end{equation}
where $F(\cdot)$ embeds the SEE defined in~\eqref{eq:see_def} through $P_{\mathrm{slot}}(t)$ from~\eqref{eq:pslot}, together with secrecy and sensing–entropy penalties, and $\mathbf{p}(t)$ denotes the Bayesian sensing posterior that couples the leader decisions to the belief-dependent constraints.
\subsection{Stackelberg Leader Control and Pricing}
\label{Stack_pricing}
At the leader layer, the IBS solves a constrained power–allocation and pricing problem that respects the power–simplex and feasibility constraints in \eqref{eq:alphabetagamma}–\eqref{eq:alphasum}. Given SEE, total transmit power, and slot–wise consumption defined in
\eqref{eq:see_def}, \eqref{eq:ptx_tot}, and \eqref{eq:pslot}, the leader's goal per slot is described by \(F(\cdot)\) in\eqref{eq:stackelberg_overall}, which maximizes SEE while penalizing secrecy-rate violations and excessive sensing entropy.
 Following that, Algorithm~\ref{Algorithm_1} implements a Bayesian–Stackelberg controller that approximates in real time. In each slot, the IBS  (i) predicts a prior AoA belief \(p_t^{-}(\theta)\) for the eavesdropper through a Gaussian convolution,  (ii) computes the predicted entropy \(H_t\) and maps it to a sensing fraction,\(\gamma_t = \Gamma(H_t)\),  (iii) updates the AN fraction \(\beta_t\) using the secrecy error \(e_t = R_s^{\text{target}} - R_s(t-1)\),  (iv) enforces the power simplex by setting \(\alpha_t = 1-\beta_t-\gamma_t\), and (v) adjusts the incentive prices \((\pi_t,\tau_t,\kappa_t)\) through clipped affine updates driven by jamming benefit, average leakage, and entropy deviation.Through this Bayesian–Stackelberg update, the IBS continuously reshapes the HN utilities and the resulting GNE so that the induced equilibrium is entropy, secrecy-aware, and explicitly oriented toward maximizing SEE, in direct correspondence with the second contribution item on adaptive BS power control.
\begin{algorithm}[t]
\caption{GNE Response and Role Switching at HNs }
\label{Algorithm_2}
    \textbf{Require:}BS splits $(\alpha_t,\beta_t,\gamma_t)$; prices $(\pi_t,\tau_t,\kappa_t)$; channels $\{h_u\}$; power limits $P_u^{\max}$; secrecy threshold $R_s^{\min}$; network-wide coupling constraints (interference, leakage).
\textbf{Ensure:}
    Equilibrium powers $p_u^\star$;  assigned roles $\text{role}_u \in \{\text{THN},\text{JHN}\}$.

\textbf{Step $1$: Initialization}\\
set $p_u^{(0)}\in[0,P_u^{\max}]$ and $\mathrm{role}_u\leftarrow \mathrm{THN}$.\;

\textbf{Step $2$: Utility}\\
Each HN solves a self-interested utility maximization from Appendix \ref{app:follower_utility_gne}.

\textbf{Step 3: Best-Response Dynamics}\;
repeat for $i=1,2,\dots$ until convergence:\;
$p_u^{(i)} \leftarrow \arg\max_{0\le p_u\le P_u^{\max}} U_u(p_u,\mathbf{p}^{(i-1)}_{-u})$ (project to satisfy $g(\mathbf{p})\le 0$ if needed).\;
stop if $\|\mathbf{p}^{(i)}-\mathbf{p}^{(i-1)}\|_2<\epsilon$.\;

\textbf{Step 4: Equilibrium Identification, GNE}\;
set $\mathbf{p}^\star\leftarrow \mathbf{p}^{(i)}$.\;

\textbf{Step 5: Role Switching Rule}\;
$\mathrm{role}_u \leftarrow \mathrm{JHN}$ if $R_{s,u}(\mathbf{p}^\star)<R_s^{\min}$, else $\mathrm{THN}$.\;

\textbf{Step 6: output}\;
forward $(\mathbf{p}^\star,\{\mathrm{role}_u\})$ to the refinement layer.\;
\end{algorithm}

\subsection{Follower-Level GNE for Power Control and Role Switching}
\label{Follower_level}

Given the leader action vector $\mathbf{a}_t$, the HNs interact on the mid-timescale through a GNE that captures power control and role selection under imperfect CSI. Thus, the composite BS-HN channel is modeled as
\begin{equation}
  \mathbf{H}_{HN}
  = \widehat{\mathbf{H}}_{HN} + \mathbf{E}_{HN}, \qquad
  \|\mathbf{E}_{HN}\|_F^2 \le \epsilon_{HN},
  \label{eq:csi_uncertainty_iBEAMS}
\end{equation}
where $\widehat{\mathbf{H}}_{HN}$ is the estimated channel, and
$\mathbf{E}_{HN}$ denotes the bounded CSI error. Under this uncertainty model, the robust secrecy rate for an active HN $u\in\mathcal{A}(t)$ is given by~\eqref{eq:secrecy_rate} as
\begin{equation}
\begin{aligned}
R_s(u,t)
= \Big[
\log_2\!\big(1+\mathrm{SINR}_{\mathrm{lu}}(u,t)\big)
\\- \max_{e \in \mathcal{E}}
\log_2\!\big(1+\mathrm{SINR}_{\mathrm{e}}(u,e,t)\big)
\Big]^+ .
\end{aligned}
\label{eq:Rs_slot_t_detailed}
\end{equation}
where $\mathrm{SINR}_{\mathrm{lu}}(u,t)$ and
$\mathrm{SINR}_{\mathrm{e}}(u,e,t)$ depend on the BS communication and AN
powers $P_{\mathrm{BS}}^{\mathrm{comm}}(t)$,
$P_{\mathrm{BS}}^{\mathrm{AN}}(t)$, the HN powers $\{P_u(t)\}$, and the
induced multi-user interference pattern.

\paragraph*{Utility Design and Secrecy–Energy–Entropy Pricing}
To align individual HN incentives with the system-level SEE objective, the local payoff for HN, $u$ at slot $t$ is developed from Appendix \ref{app:follower_utility_gne}, where the utility design can be
\begin{equation}
\begin{aligned}
U\!\big(P_u(t),\text{role}_u(t); \mathbf{a}_t,\mathbf{P}_{-u}(t)\big)
   &= \eta_u R_s(u,t) - c_u P_u(t)  \\
   &\quad -\, \tau_t \Xi_u(t) + \pi_t J_u(t) + \kappa_t I(t)
\end{aligned}
\label{eq:HN_utility_detailed}
\end{equation}

In general, this utility $U$, can be viewed as a cost-per-node SEE measure, incorporating the benefit of secrecy-rate, the cost of power, the penalty for leakage, the reward for jamming and the gain of information based on shared sensing.
In this way, the leader’s prices $(\pi_t,\tau_t,\kappa_t)$ implement the “Secrecy–Energy–Entropy-aware pricing’’ mechanism from the iBEAMS contributions, embedding ISAC BS priorities directly into follower payoffs.
\paragraph*{Feasible Action Sets and Coupling Constraints}
Each HN’s power decision is constrained by both local hardware limits and network-wide coupling conditions. These constraints can be expressed as
\begin{align}
  0 \le P_u(t) &\le P_u^{\max}, \qquad \forall u,
  \label{eq:HN_power_bound}\\
  \sum_{u\in\mathcal{A}(t)} P_u(t)
  &\le P_{\mathrm{FJ}}^{\max},
  \label{eq:FJ_total_budget}\\
  \Xi_u(t) &\le \Xi_{\max}, \qquad \forall u,
  \label{eq:leak_constraint_detailed}\\
  g\big(P_1(t),\dots,P_u(t)\big) &\le 0,
  \label{eq:global_coupling_constraint}
\end{align}
where \eqref{eq:HN_power_bound} enforces per-node power bounds,
\eqref{eq:FJ_total_budget} limits the aggregate friendly-jamming power,\eqref{eq:leak_constraint_detailed} limits the secrecy leakage for each HN, and \eqref{eq:global_coupling_constraint} compactly represents additional  interference or SINR feasibility constraints.

Collecting these conditions, the global feasible set of power profiles are created as
\begin{equation}
\mathcal{P}(t)
=\Big\{
\mathbf{P}(t) = (P_1(t),\dots,P_u(t))
\;\Big|\;
\eqref{eq:HN_power_bound}–\eqref{eq:global_coupling_constraint}\ \text{combined}\Big\}.
\label{eq:global_feasible_set}
\end{equation}
For a given HN $u$, its admissible decisions are then given by the projection of $\mathcal{P}(t)$ onto the $u$-th coordinate:
\begin{equation}
\mathcal{P}_u(t)
=\Big\{
P_u(t)\;\Big|\;
\exists\,\mathbf{P}_{-u}(t)\ \text{such that}\
\big(P_u(t),\mathbf{P}_{-u}(t)\big)\in\mathcal{P}(t)
\Big\}.
\label{eq:Pk_feasible_set}
\end{equation}
\paragraph*{GNE Formulation and Best-Response Approximation}
Since utility, $U$ depends on all other powers through the secrecy rate, leakage, and jamming terms, the power profile $\{P_u^\star(t)\}_u$ is a GNE if
\begin{equation}
\begin{aligned}
U\!\big(P_u^\star(t), P_{-u}^\star(t)\big)
   &\ge U_u\!\big(P_u(t), P_{-u}^\star(t)\big), \\
&\forall\, P_u(t)\in\mathcal{P}_u(t),\ \forall\, u .
\end{aligned}
\label{eq:GNE_definition}
\end{equation}
taken together from \ref{app:follower_utility_gne}, that says no HN can unilaterally increase its utility while respecting all local and shared constraints. The Algorithm~\ref{Algorithm_2} implements an iterative best-response scheme to approximate this GNE on a discrete power grid. Starting from an initial power profile $\{P_u^{(0)}(t)\}_u$ , where all nodes are provisionally treated as THNs, each HN repeatedly solves
\begin{equation}
  P_u^{(i)}(t)
  = \arg\max_{P_u(t)\in\mathcal{P}_u(t)}
  U\big(P_u(t),P_{-u}^{(i-1)}(t)\big),
  \label{eq:BR_update}
\end{equation}
until the updates satisfy
$\|P_u^{(i)}(t)-P_u^{(i-1)}(t)\|\le\epsilon$ for all $k$. The resulting fixed point $\{P_u^\star(t)\}_k$ provides an implementable approximation to the GNE defined in~\eqref{eq:GNE_definition}.

\paragraph*{Secrecy-Based Role Switching and Energy-Aware Jamming}
Once the equilibrium powers are obtained, iBEAMS maps each HN to a transmitting or jamming role using the secrecy-based rule
\begin{equation}
  \text{role}_u(t)
  =\begin{cases}
    \text{JHN}, & R_s(u,t) < R_s^{\min},\\[0.5ex]
    \text{THN}, & \text{otherwise},
  \end{cases}
  \label{eq:role_switch_rule}
\end{equation}
where $R_s^{\min}$ is the minimum acceptable secrecy rate for a legitimate transmission. HNs that cannot satisfy $R_s^{\min}$ under the current equilibrium are reconfigured as friendly jammers, thereby contributing additional interference toward eavesdroppers while incurring power and leakage-related penalties in their utilities. The resulting equilibrium powers and roles, received from Algorithm~\ref{Algorithm_2} to the Bayesian refinement layer (Algorithm~\ref{algorithm_3}), determine the strategic HN role switching, and energy-aware jamming contribution within a rigorous GNE framework, and couple naturally with the leader’s entropy, and secrecy-aware pricing policy.
\begin{algorithm}[t]
\caption{Bayesian Cooperative Refinement and Posterior-Aligned Jamming}
\label{algorithm_3}
   \textbf{Require:} THN/JHN sets from GNE; equilibrium powers $p_k^\star$;
    predicted posterior $p_t^{-}$; sensing measurements $\{z(\theta)\}$;  BS beam shapes (AN/data).
    \textbf{Ensure:}Updated posterior $p_t$;refined jamming powers;posterior aligned jamming field with THN-protective nulls.

\textbf{Step $1$: Coalition Formation} form jamming coalitions among JHNs near dominant peaks of $p_t^{-}$; form protection sets for THNs.

\textbf{Step $2$: Posterior Likelihood Update}\\Compute pseudo-likelihood from \eqref{eq:likelihood_update}; update from the equation \eqref{eq:posterior_update}; 
normalize $\int p_t(\theta)d\theta=1$.\;

\textbf{Step $3$: Cooperative Jamming Optimization}\\
For each jamming coalition $\mathcal{C}$, update $\{p_k\}_{k\in\mathcal{C}}$ via \eqref{eq:coalition_refine_main} and local iterative improvement that increases $\sum_{u\in\mathrm{THN}}R_{s,u}$ subject to $0\le p_k\le P_k^{\max}$ and THN leakage constraints.

\textbf{Step $4$: Posterior-Aligned Beam/Jamming Field Synthesis}\\
Construct AN lobes aligned with $\arg\max p_t(\theta)$, Insert deep nulls toward THNs or protected angles, Adjust jamming phases coherently within each coalition.

\textbf{ Step $5$: Stop}\\
terminate if $\Delta R_s^{\mathrm{sum}}<\delta$; output $(p_t,\{p_k\},\text{jamming field})$.\;
\end{algorithm}
\begin{table}[t]
\centering
\caption{Key simulation parameters}
\label{tab:sim_params_ibeams_2col}
\renewcommand{\arraystretch}{1.02}
\setlength{\tabcolsep}{4pt}
\begin{tabularx}{\linewidth}{@{} X l @{}}
\toprule
\textbf{Parameter} & \textbf{Value} \\
\midrule

\multicolumn{2}{@{}l@{}}{\textbf{\underline{\makebox[\linewidth][l]{Carrier, bandwidth, and noise}}}}\\[-1pt]
Carrier frequency, $f_c$ & $\SI{28}{GHz}$ to $\SI{3}{THz}$ \\
System bandwidth, $BW$ & $\SI{100}{MHz}$ \\
Thermal noise PSD, $N_0$ (at $\approx\SI{290}{K}$) & $-174~\si{dBm/Hz}$ \\
Receiver noise figure, NF & $\SI{7}{dB}$ \\[-0.5pt]

\multicolumn{2}{@{}l@{}}{\textbf{\underline{\makebox[\linewidth][l]{ISAC base station (IBS) array and geometry}}}}\\[-1pt]
ISAC BS ULA antennas, $N_t$ & $128$ \\
IBS position, $\mathbf{p}_{\mathrm{BS}}$ (m) & $[0,\,0,\,10]^\top$ \\ [-0.5pt]

\multicolumn{2}{@{}l@{}}{\textbf{\underline{\makebox[\linewidth][l]{Hybrid edge node (HN) array and geometry}}}}\\[-1pt]
Number of HNs, $K_{\mathrm{HN}}$ & $25$ \\
HN array size, $N_{t,\mathrm{HN}}$ & $4\times4 = 16$ \\[-0.5pt]

\multicolumn{2}{@{}l@{}}{\textbf{\underline{\makebox[\linewidth][l]{Large-scale and small-scale channel model}}}}\\[-1pt]
Path-loss exponent, $n_{\mathrm{PL}}$ & $2.2$ \\
Shadowing, $\sigma_{\mathrm{sh}}$ & $\SI{3}{dB}$ \\
Rician $K$-factor, $K_{\mathrm{Ric}}$ & $\SI{10}{dB}$ \\[-0.5pt]

\multicolumn{2}{@{}l@{}}{\textbf{\underline{\makebox[\linewidth][l]{Power budgets}}}}\\[-1pt]
Max BS power, $P_{\mathrm{BS}}^{\max}$ & $\SI{20}{W}$ / $\SI{43}{dBm}$ \\
Initial BS power, $P_{\mathrm{BS}}^{(0)}$ & $\SI{15}{W}$ / $\SI{41.76}{dBm}$ \\
Max per-HN power, $P_{\mathrm{HN}}^{\max}$ & $\SI{1.5}{W}$ / $\SI{31.76}{dBm}$ \\[-0.5pt]

\multicolumn{2}{@{}l@{}}{\textbf{\underline{\makebox[\linewidth][l]{Leader (Stackelberg) variables and bounds}}}}\\[-1pt]
BS power split $(\alpha,\beta,\gamma)$ (data/AN/sensing) & $(0.6,\,0.2,\,0.2)$ \\
Incentive prices $(\pi_t,\tau_t,\kappa_t)$ (reward/penalty/info) & $(0.7,\,0.3,\,0.1)$ \\
Bounds on $\pi_t$, $(\pi_{\min},\pi_{\max})$ & $(0,1)$ \\
Bounds on $\tau_t$, $(\tau_{\min},\tau_{\max})$ & $(0,1)$ \\
Bounds on $\kappa_t$, $(\kappa_{\min},\kappa_{\max})$ & $(0,1)$ \\[-0.5pt]

\multicolumn{2}{@{}l@{}}{\textbf{\underline{\makebox[\linewidth][l]{Eve AoA posterior tracking (Bayesian layer)}}}}\\[-1pt]
AoA grid, $\theta$ & $[-90,90]^\circ$ \\
Initial kernel width, $\sigma_{\mathrm{ker}}^{(0)}$ & $10^\circ$ \\
AoA measurement noise std.\ dev., $\sigma_{\mathrm{meas}}$ & $5^\circ$ \\[-0.5pt]

\multicolumn{2}{@{}l@{}}{\textbf{\underline{\makebox[\linewidth][l]{GNE / best-response settings}}}}\\[-1pt]
GNE tolerance, $\varepsilon_{\mathrm{GNE}}$ & $10^{-3}$ \\
Max Best-Response (BR) iterations per slot, $I_{\max}$ & $50$ \\
\bottomrule
\end{tabularx}
\end{table}
\subsection{Bayesian Sensing, Entropy Constraint, and Geometry-Aware Refinement}

The Bayesian layer addresses the sensing-entropy and belief-coupled sensing problem in Section~\ref{sensing entropy_problem}. The IBS maintains the posterior $\mathbf{p}(t)$ as in \eqref{eq:belief_vector_slot_t}, with entropy
\eqref{eq:entropy_slot_t} constrained by \ref{eq:entropy_constraint} and the penalty by \ref{eq:entropy_penalty}, which appears in the leader objective and the sensing subproblem.

Algorithm~\ref{Algorithm_1} first predicts the AoA belief through
\begin{equation}
  p_t^{-}(\theta)
  = \big(K(\sigma_t) * p_{t-1}\big)(\theta),
  \label{eq:belief_prediction}
\end{equation}
where $K(\sigma_t)$ is a Gaussian kernel with bandwidth $\sigma_t$ modeling mobility and temporal evolution. Therefore, the predicted entropy
\begin{equation}
  H_t=  -\int p_t^{-}(\theta)\log_2 p_t^{-}(\theta)\,d\theta
  \label{eq:entropy_predicted}
\end{equation}
is mapped to a sensing power fraction through $\gamma_t = \Gamma(H_t)$ with
$\Gamma'(\cdot)>0$, $\Gamma(0)=\gamma_{\min}$, and
$\Gamma(H^{\max})=\gamma_{\max}$ so that more power is devoted to sensing when uncertainty is high.

Given $p_t^{-}$ and sensing measurements $\{z(\theta)\}$,
Algorithm~\ref{algorithm_3} forms a pseudo-likelihood
\begin{equation}
  L(\theta) = z(\theta)^{k_{\mathrm{eff}}},
  \label{eq:likelihood_update}
\end{equation}
and computes the posterior as
\begin{equation}
  p_t(\theta) \propto p_t^{-}(\theta)\,L(\theta),
  \label{eq:posterior_update}
\end{equation}
normalized to satisfy \eqref{eq:belief_vector_slot_t}. The updated posterior feeds the Bayesian refinement mechanisms. This refinement mechanism, the updated posterior belief $p_t(\theta)$, subsequently drives three Bayesian refinement mechanisms that couple sensing, cooperation, and jamming. 

Firstly, it induces belief-aware coalition formation: JHNs whose angular sectors overlap with dominant posterior peaks, for example large $p_t(\theta)$) self-organize into jamming coalitions, while THNs with similar QoS and secrecy requirements form protection coalitions, thereby explicitly linking cooperative behavior to the belief state. Secondly, within each jamming coalition $\mathcal{C}$, the cooperative jamming powers $\{P_k(t)\}_{k\in\mathcal{C}}$ are refined by conceptually maximizing the aggregate secrecy of the served THNs, subject to per-node power limits and bounded leakage.
\begin{equation}
\begin{aligned}
\max_{\{P_u(t)\le P_u^{\max}\}_{u\in\mathcal{C}}}\;
    &\sum_{u\in\mathrm{THN}} R_s(u,t) \\
\text{s.t.}\quad
    &\Xi_u(t) \le \Xi_{\max},\ \forall u\in\mathrm{THN}.
\end{aligned}
\label{eq:coalition_opt}
\end{equation}
where the AN and jamming beams are shaped according to $p_t(\theta)$ to concentrate energy along secrecy-critical directions while limiting leakage toward protected THNs. Third, the posterior is leveraged for posterior-aligned beam synthesis, the IBS steers AN/jamming beams toward $\arg\max_{\theta} p_t(\theta)$ and enforces nulls (or reduced gain) in directions associated with protected THNs. Moreover, the phases of JHNs within each coalition are adjusted coherently so that their contributions add constructively at the eavesdropper and destructively at legitimate receivers. In addition, the kernel bandwidth controlling belief prediction is adaptively updated as
\begin{equation}
  \sigma_{t+1}
  =\sigma_t + \eta_\sigma\big(H_t - H^{\max}\big),
  \label{eq:kernel_update}
\end{equation}
which broadens the prediction when uncertainty is higher than desired and narrows it when the posterior is already sharp. Collectively, these mechanisms instantiate the proposed Bayesian cooperative refinement by ensuring that both sensing effort and cooperative jamming geometry are directly governed by the belief-entropy state.
\subsection{Cooperative Jamming Refinement (Layer 3)}
Given the updated posterior $p_t(\theta)$ and the provisional THN/JHN sets from Layer~2, iBEAMS performs a coalition-based refinement in which each jamming coalition $\mathcal{C}\subseteq\mathcal{J}(t)$ updates only its local jamming powers while keeping non-coalition actions fixed. Concretely, the coalition powers are refined by maximizing the aggregate secrecy of the active THNs,
\begin{equation}
\begin{aligned}
\{P_k^{\mathrm{new}}(t)\}_{k\in\mathcal{C}}
&=\arg\max_{\{0\le P_k(t)\le P_k^{\max}\}_{k\in\mathcal{C}}} \\
&\quad \sum_{u\in\mathcal{T}(t)}
R_s\!\big(u,t;\mathbf{p}_{\mathcal{C}}(t),\mathbf{p}_{-\mathcal{C}}(t)\big),
\end{aligned}
\label{eq:coalition_refine_main}
\end{equation}
subject to THN-protection (bounded leakage/interference toward legitimate users) and posterior-weighted shaping constraints that concentrate jamming energy along high-probability AoA directions and enforce nulls toward protected THNs. The detailed derivation and constraint instantiation of \eqref{eq:coalition_refine_main} are provided in Appendix~\ref{app:coop_jam_derivation}.
\subsection{Utility design of the three layer}
The slot-$t$ decision tuple is defined as \(
\mathbf{x}_t \triangleq \big(\mathbf{a}_t,\mathbf{z}_t,\mathbf{b}_t\big)
\), and the overall utility is defined as $\mathcal{J}_t(\mathbf{x}_t)
\triangleq\;$, with the detailed formulation described in Appendix \ref{app:unified_three_layer_utility}. 

Taken together, the three layers of the iBEAMS framework jointly solve the coupled optimization problems of Section~\ref{problemform} within a single hierarchical game structure. At the top, the Stackelberg leader layer realizes secrecy, and entropy-aware BS power control and power splitting through the constrained optimization with the SEE-based objective in \eqref{eq:stackelberg_overall}, thereby enforcing the power-simplex, secrecy, and sensing-entropy constraints through explicit penalty terms. On the follower side, the GNE layer leverages the utility definition in \eqref{eq:HN_utility_detailed} together with the feasibility conditions in \eqref{eq:HN_power_bound}–\eqref{eq:global_coupling_constraint} to embed energy cost, secrecy gain, leakage, and information gain into a unified payoff, and computes equilibrium transmit powers and roles using the best-response and role-switching rules in \eqref{eq:BR_update}–\eqref{eq:role_switch_rule}. Finally, the Bayesian layer evolves the belief state through \eqref{eq:belief_prediction}–\eqref{eq:posterior_update}, and coalition-level refinement in \eqref{eq:coalition_opt}, while shaping sensing beams and cooperative jamming fields according to the inferred geometry of the eavesdropper. 
In the next section, we evaluate the performance of this integrated iBEAMS framework through detailed numerical simulations and compare its secrecy, energy-efficiency, and sensing behavior against relevant baselines.

\begin{figure*}[t]
\centering
\subfloat[]{\includegraphics[width=0.35\textwidth]{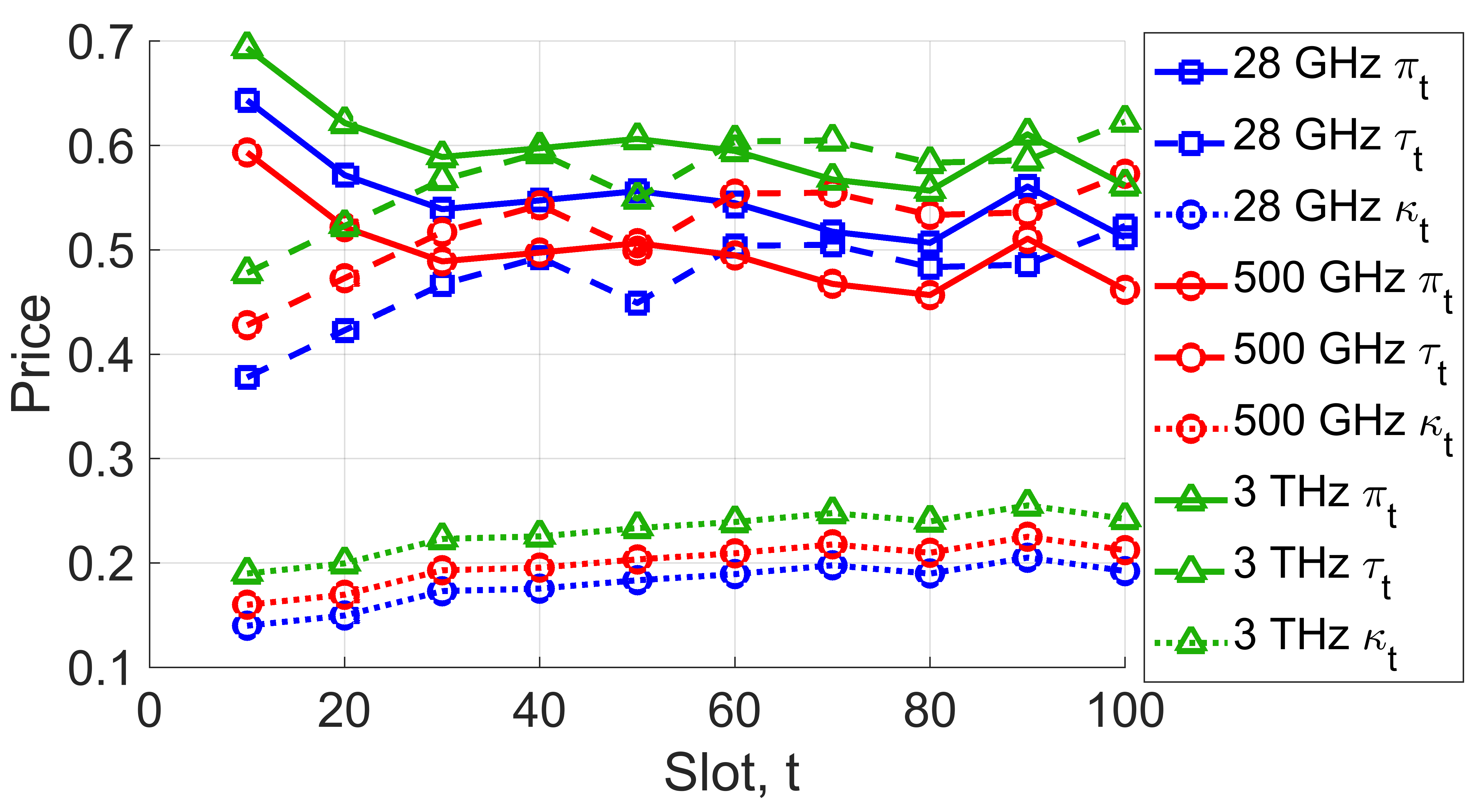}\label{fig:alg1_prices}}
\hfill
\subfloat[]{\includegraphics[width=0.35\textwidth]{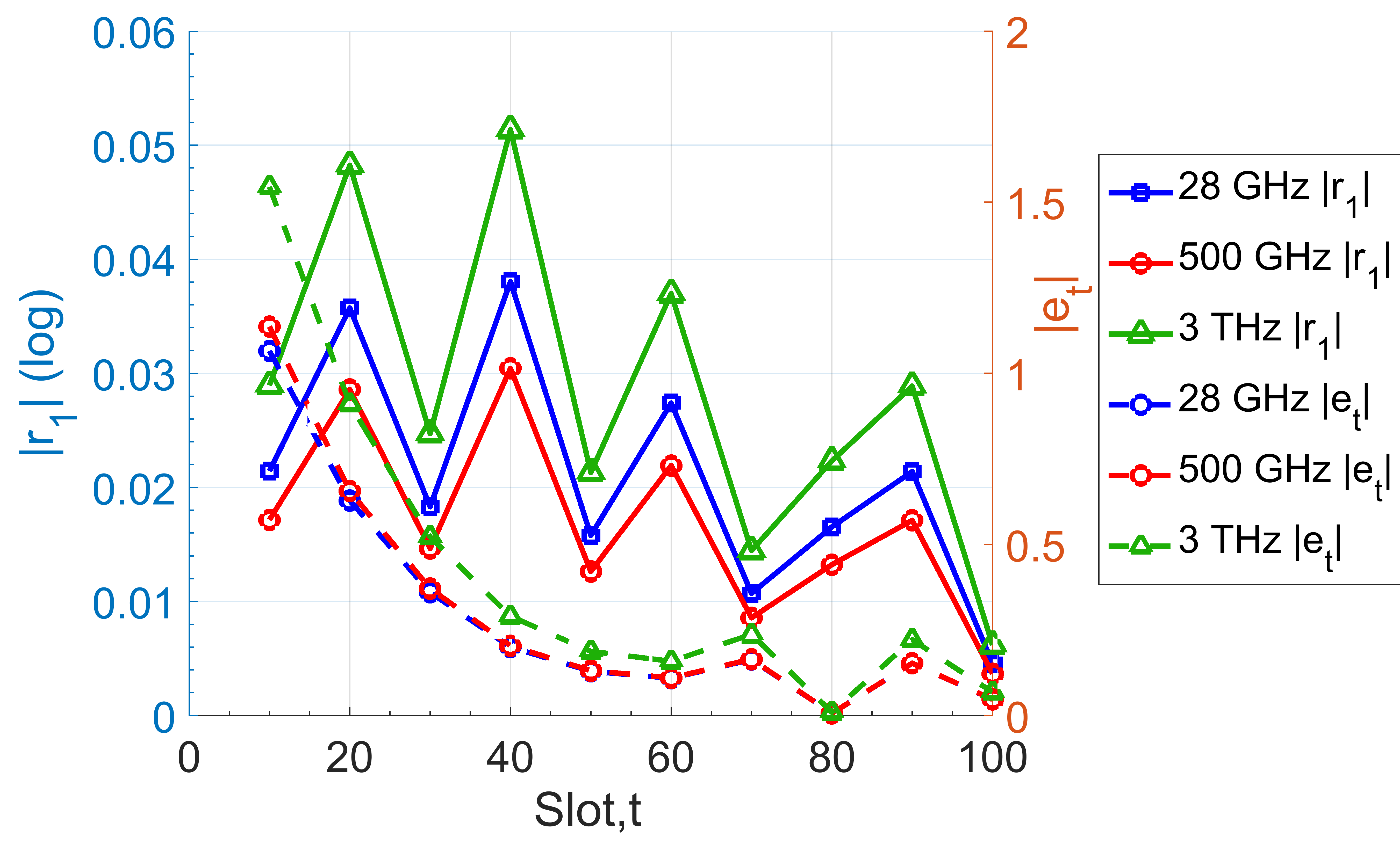}\label{fig:alg1_secerror}}
\hfill
\subfloat[]{\includegraphics[width=0.25\textwidth]{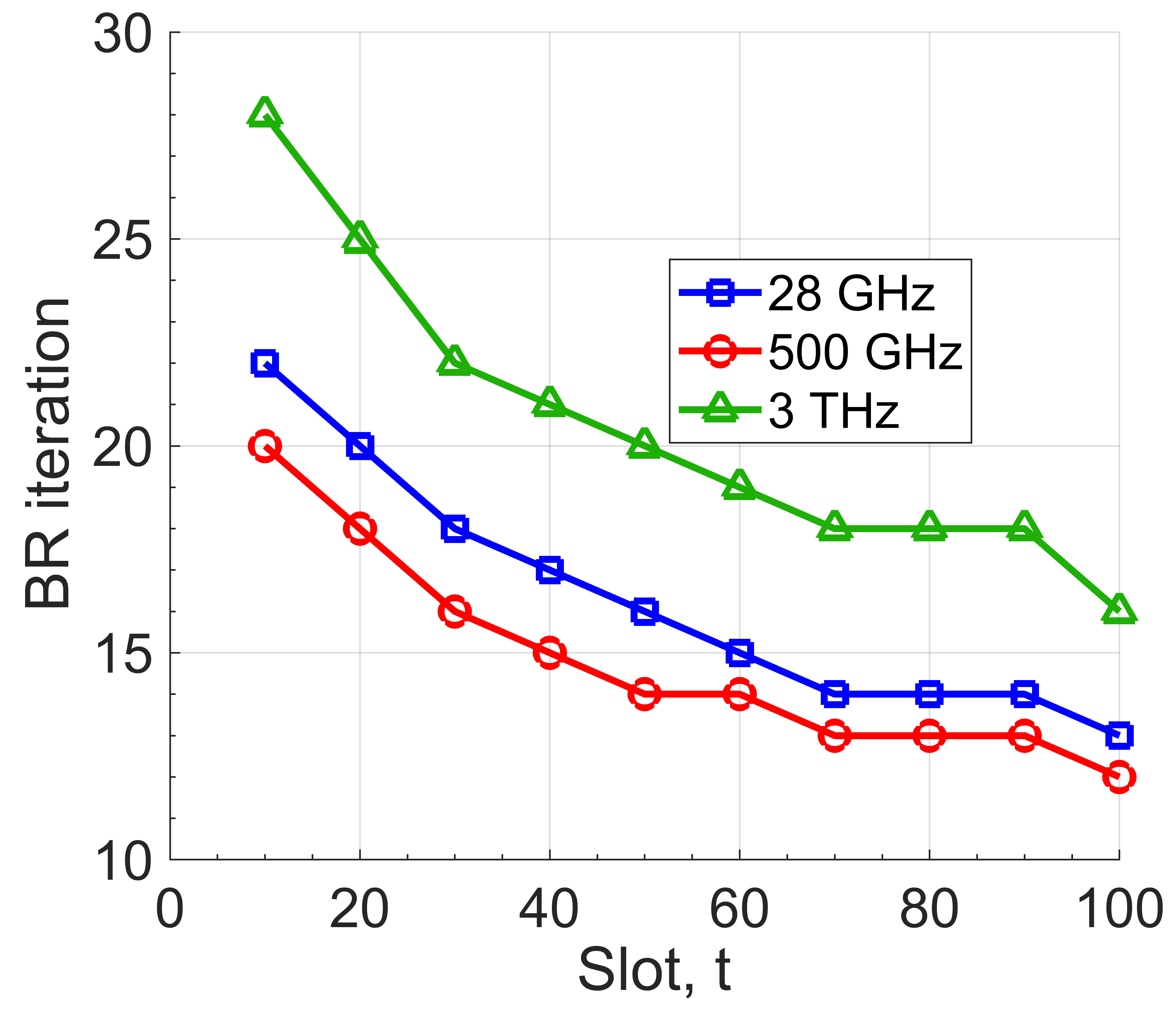}\label{fig:alg2_BR iterations}}
\vspace{-0.4cm}
\subfloat[]{\includegraphics[width=0.325\textwidth]{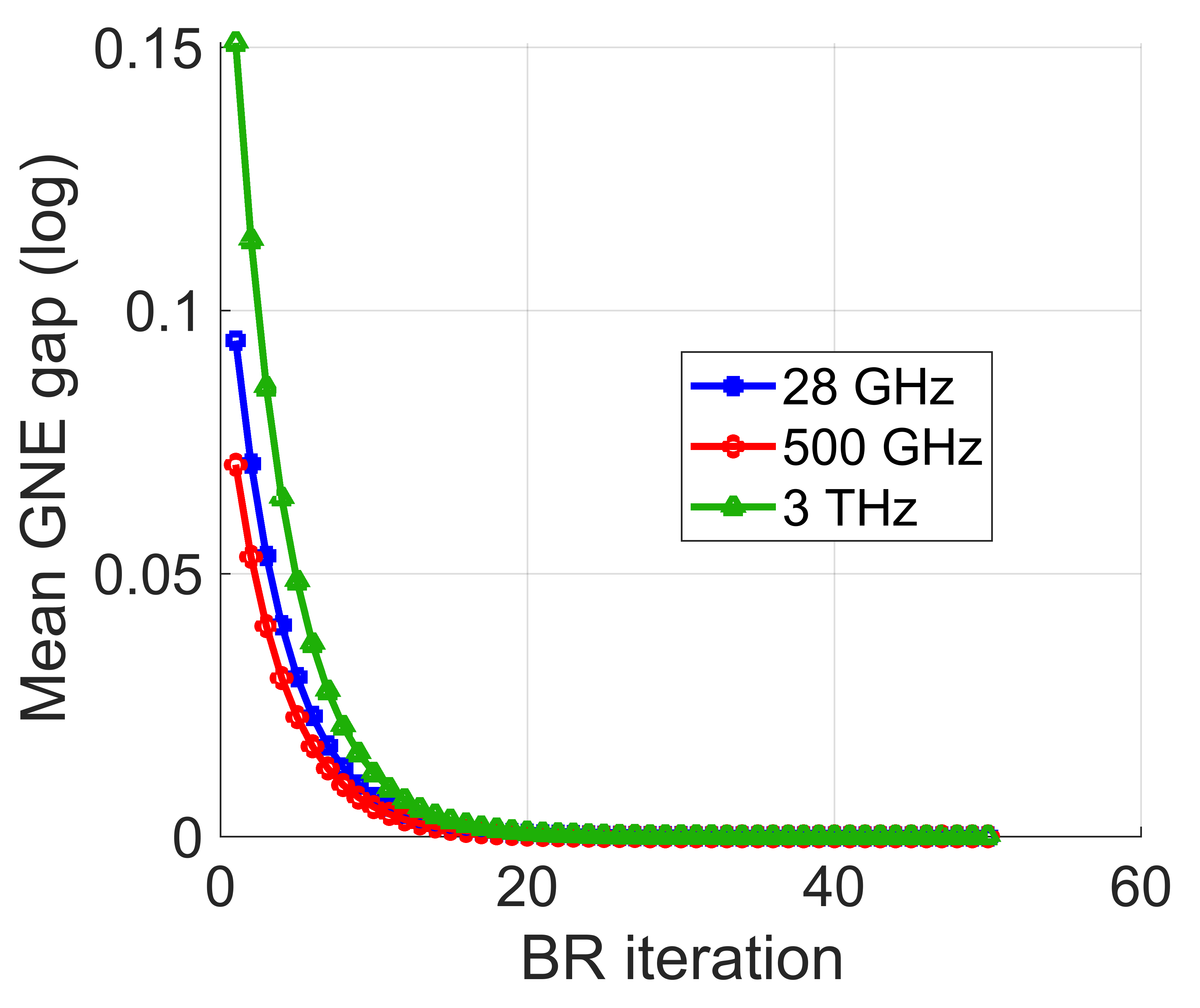}\label{fig:alg2_gnegap}}
\hfill
\subfloat[]{\includegraphics[width=0.325\textwidth]{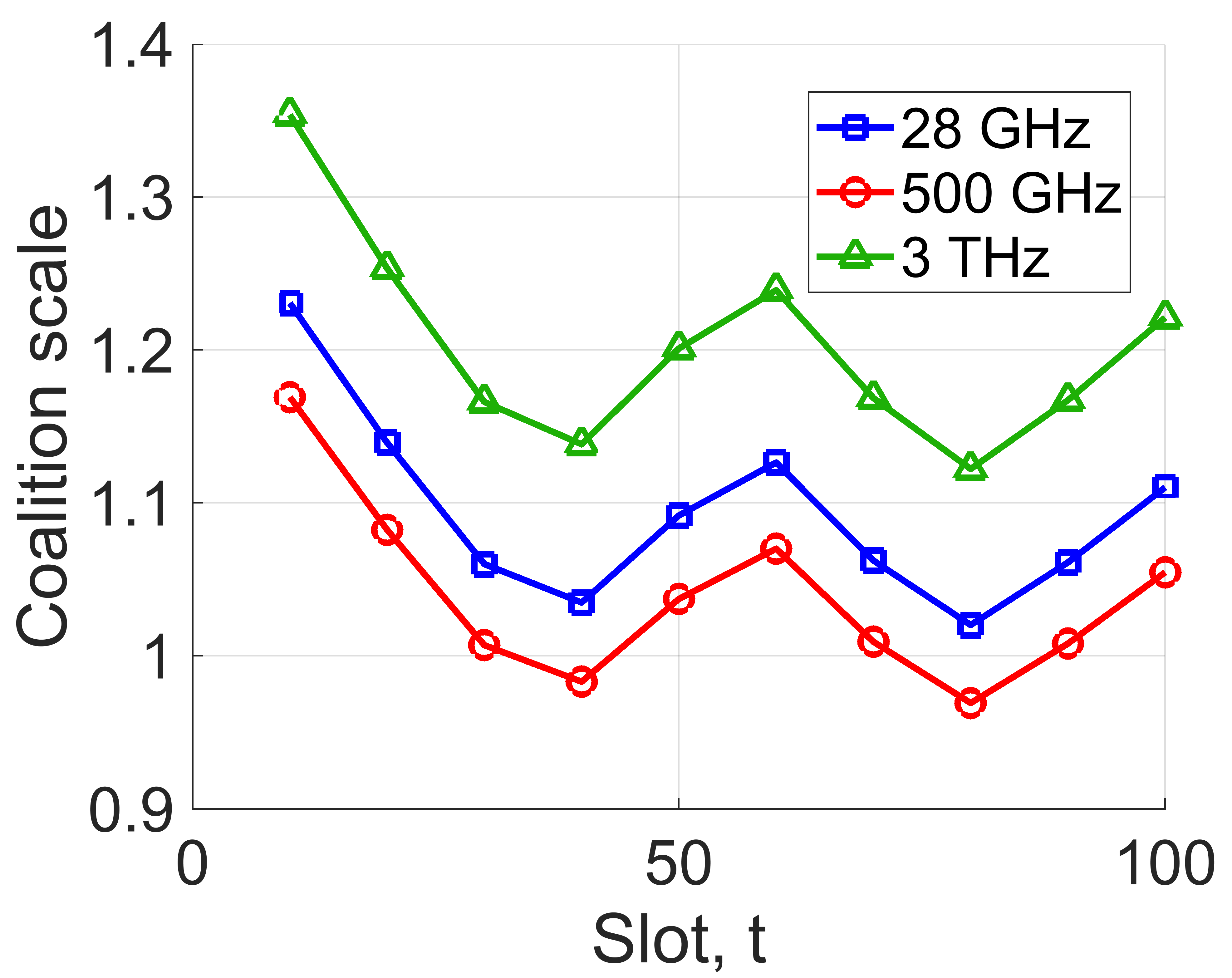}\label{fig:alg3_coalition}}
\hfill
\subfloat[]{\includegraphics[width=0.33\textwidth]{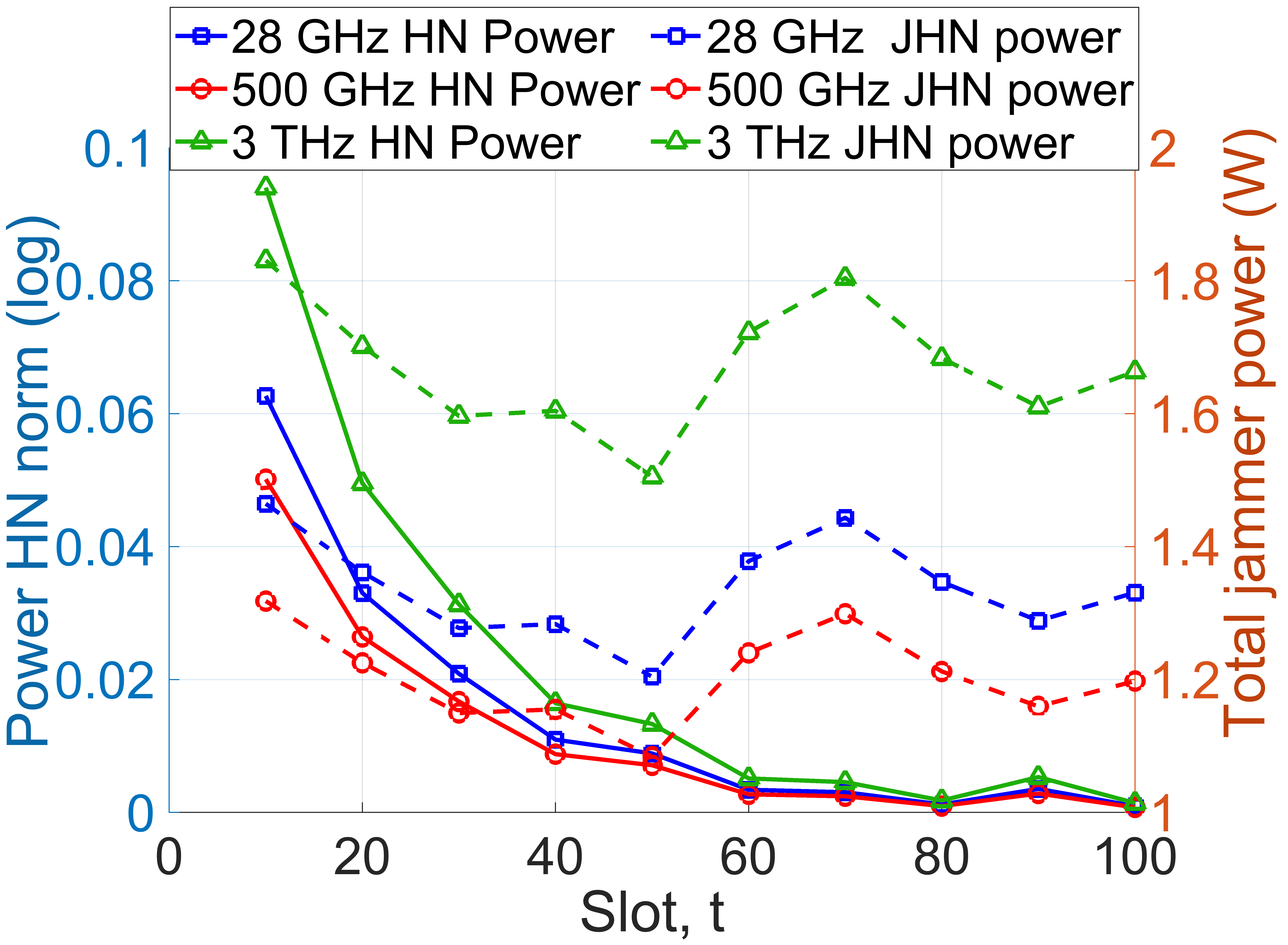}\label{fig:alg3_power}}
\caption{Convergence behavior of the proposed iBEAMS three-layer optimization of Algorithms~1-3 across frequency ($28GHz$, $500GHz$, and $3THz$). Algorithm~\ref{Algorithm_1}  (Leader) convergence in terms of intensive prices \ref{fig:alg1_prices} and leader residual, $|r1|$, and secrecy error $|e_t|$, \ref{fig:alg1_secerror}. Algorithm~\ref{Algorithm_2} GNE convergence in terms of Best Response (BR) iteration with GNE gap, \ref{fig:alg2_BR iterations} and \ref{fig:alg2_gnegap}. Finally, Algorithm~\ref{algorithm_3} Bayesian-Cooperative refinement convergence with coalition \ref{fig:alg3_coalition}, and HN power Vs. Jammer power comparison \ref{fig:alg3_power}}
\label{fig:alg_conv}
\end{figure*}

\section{Performance Evaluation}
\label{num_eval}

\subsection{Simulation Setup}
\label{simulation}

We evaluate the proposed iBEAMS framework in a downlink ISAC cellular scenario spanning multiple carrier frequencies ($28 GHZ - 3THZ$), while maintaining a fixed system bandwidth, as summarized in Table~\ref{tab:sim_params_ibeams_2col}. 
The ISAC base station (IBS) is modeled as ULA with half-wavelength spacing and a fixed location, and a set of HNs are considred with compact planar antenna arrays to support both data transmission and cooperative jamming. The propagation environment follows a distance-dependent large-scale loss model with log-normal shadowing and Rician small-scale fading, using the channel parameters in Table~\ref{tab:sim_params_ibeams_2col}. The receiver noise is computed from the PSD thermal noise and noise figure listed in Table~\ref{tab:sim_params_ibeams_2col}. 
The transmission power is limited by the IBS and per-HN power budgets in Table~\ref{tab:sim_params_ibeams_2col}. At each slot, the IBS (leader) updates the power-splitting variables and Stackelberg incentive prices within the specified bounds (Table~\ref{tab:sim_params_ibeams_2col}), and the HNs (followers) update their strategies through a per-slot GNE best-response routine using the convergence tolerance and iteration cap in Table~\ref{tab:sim_params_ibeams_2col}. In parallel, the Bayesian layer maintains a belief over the eavesdropper AoA on the discretized angular grid and refines it through a kernel-based prediction and measurement update, with kernel/measurement settings as given in Table~\ref{tab:sim_params_ibeams_2col}.
\subsection{Numerical Analysis}
\vspace{-0.15cm}
\label{num_analysis}
In this section, we discuss the simulation results to evaluate the proposed approach of iBEAMS unified framework. 

To begin with, Figures~\ref{fig:alg_conv} validates the convergence of the proposed three-layer iBEAMS control architecture across $28$~GHz, $500$~GHz, and $3$~THz.
At the leader layer (Algorithm~\ref{Algorithm_1}, Fig.~\ref{fig:alg1_prices} shows that the BS control variables converge in the sense of diminishing update magnitudes and sustained objective tracking: the power-split triplet $(\alpha_t,\beta_t,\gamma_t)$ and the Stackelberg prices $(\pi_t,\tau_t,\kappa_t)$ undergo an initial transient and then evolve around stable operating points depending on ISAC BS Bayesian-Stackelberg Leader. Simultaneously, the secrecy tracking error $|e_t|$ decreases toward a small steady regime while the leader (ISAC BS) residual $|r_1|$ remains bounded, indicating stable closed-loop behavior under time-varying fading/mobility effects (\ref{fig:alg1_secerror}). 

\begin{figure*}[t]
\centering
\subfloat[]{%
\includegraphics[width=0.23\textwidth]{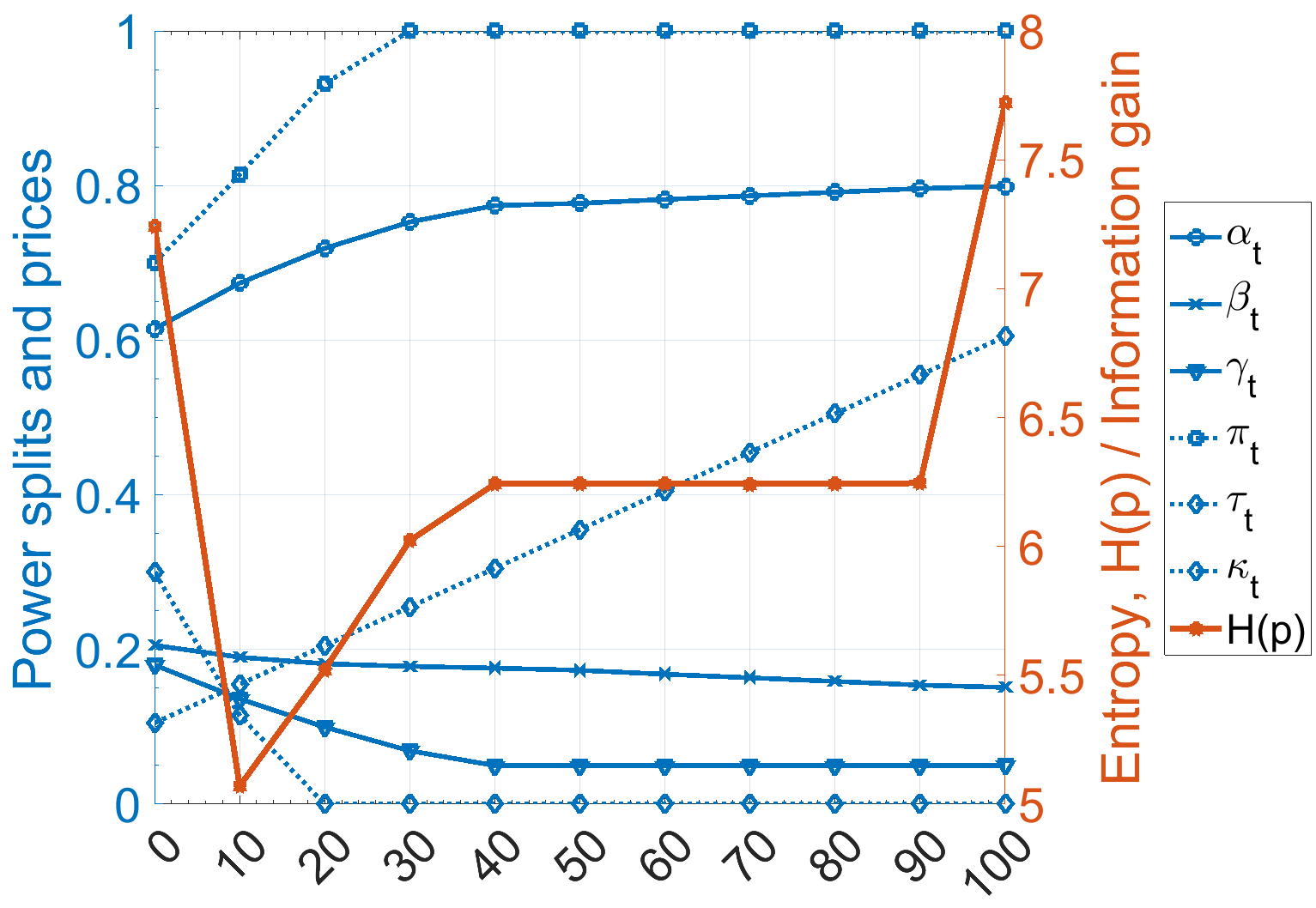}%
\label{fig:powerfraction_entropy}}%
\hfill
\subfloat[]{%
\includegraphics[width=0.23\textwidth]{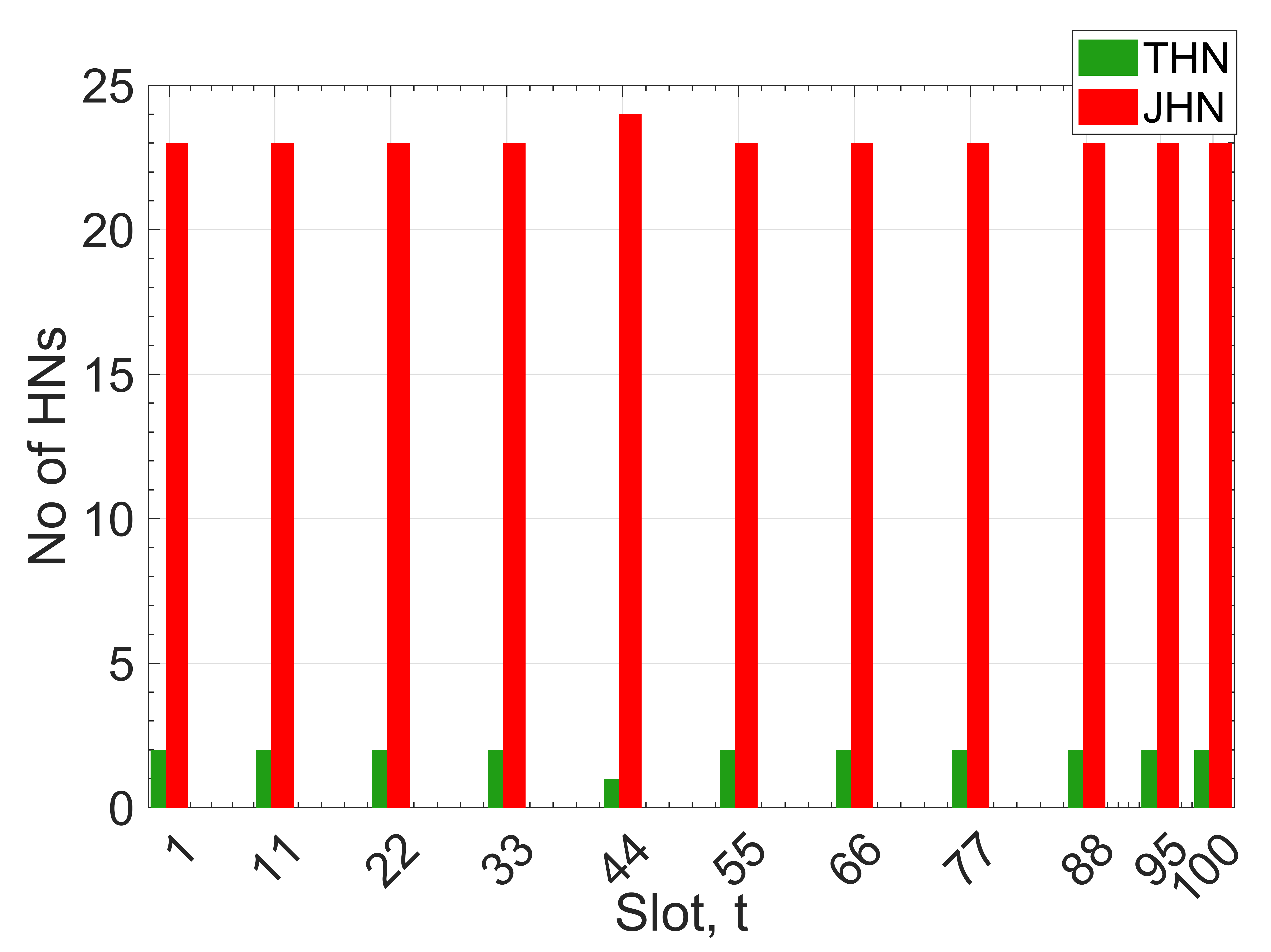}%
\label{fig:roleswitch_bar}}%
\hfill
\subfloat[]{%
\includegraphics[width=0.23\textwidth]{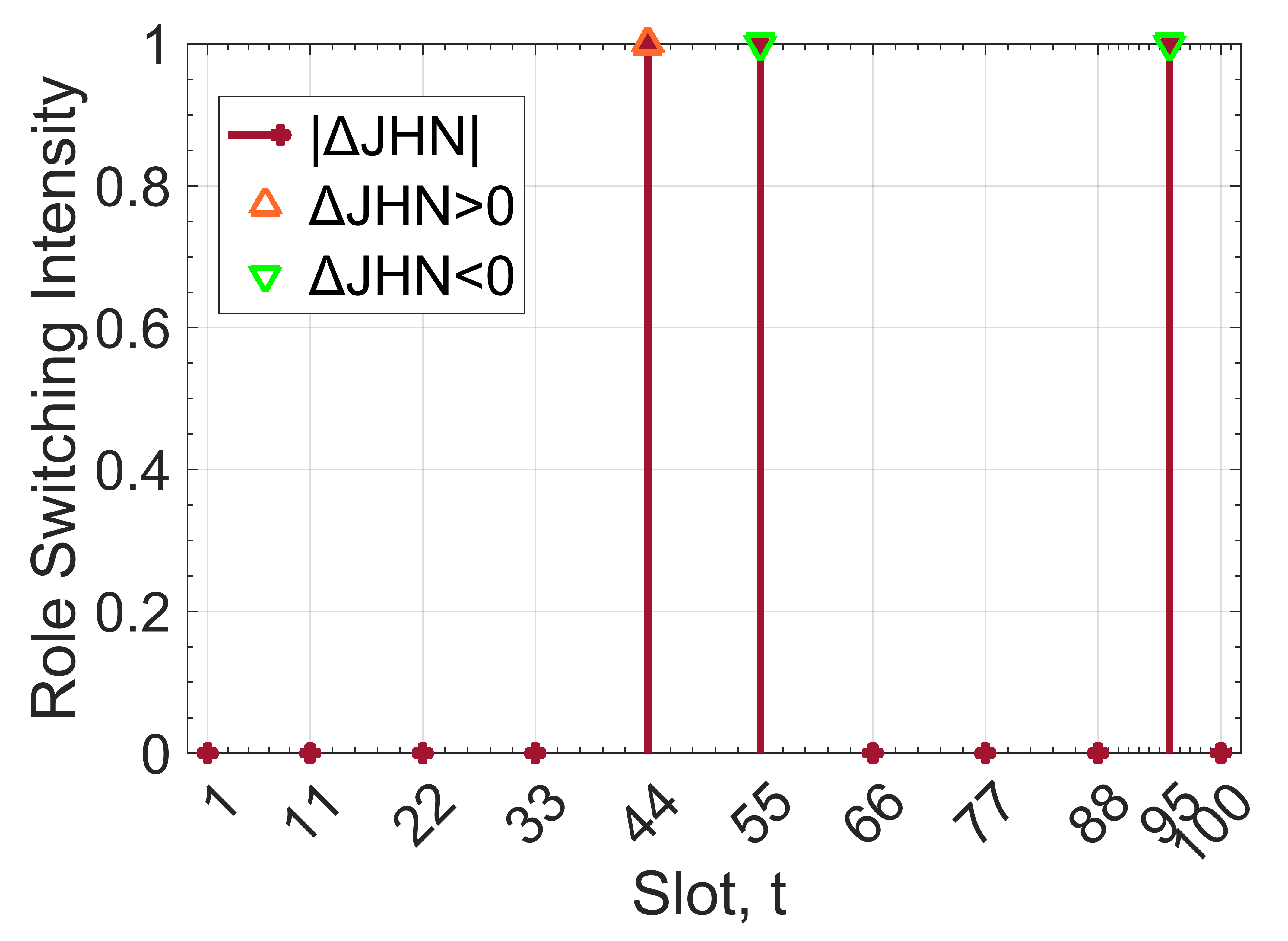}%
\label{fig:roleswitch_intensity}}%
\hfill
\subfloat[]{%
\includegraphics[width=0.23\textwidth]{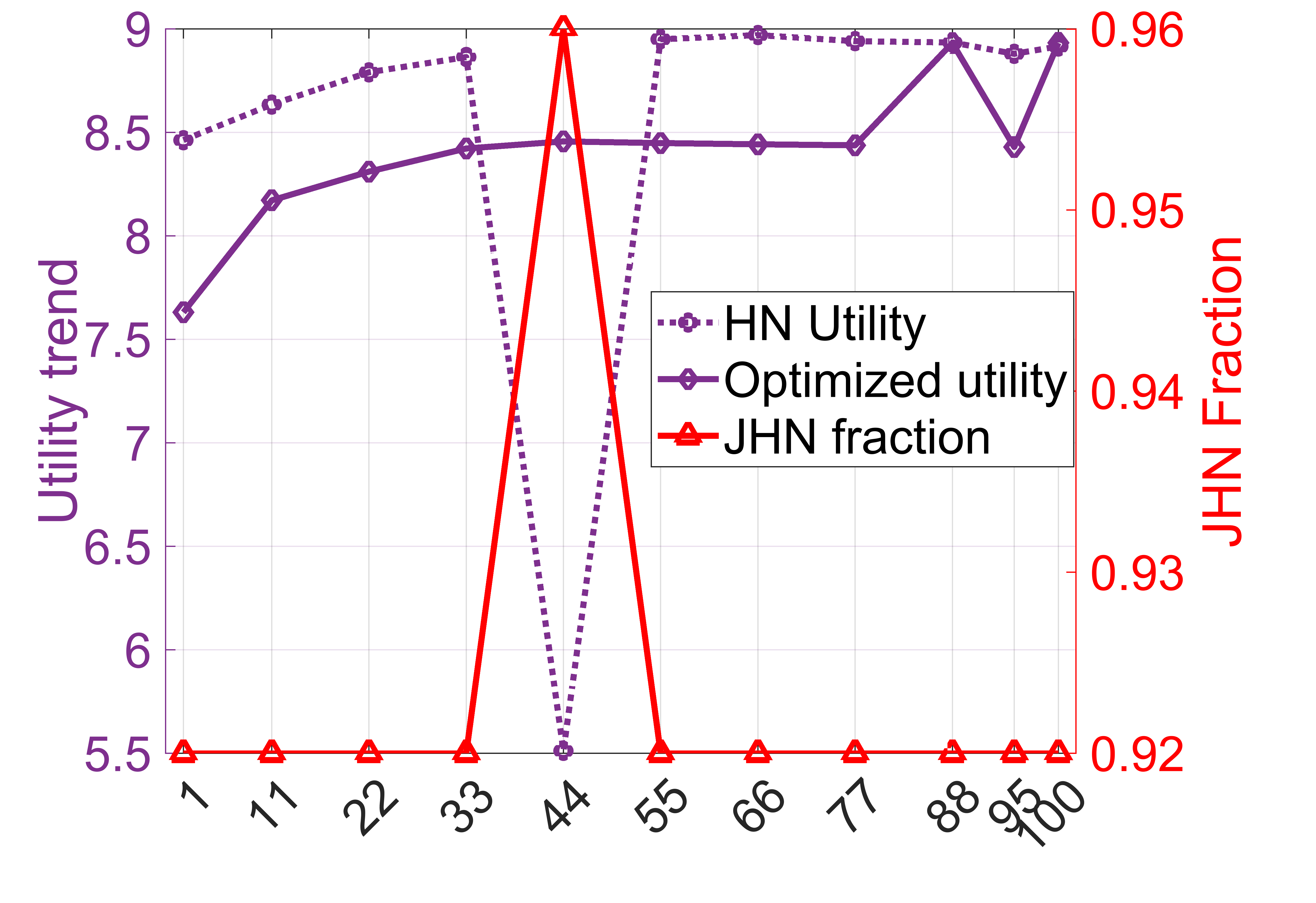}%
\label{fig:roleswitch_utility}}%
\vspace{-0.2cm}
\caption{ iBEAMS framework's performance at $28$~GHz: (a) power fractions and entropy trend, (b) THN/JHN role composition, (c) role-switching intensity, and (d) utility and jammer fraction.}
\label{fig:ibeams at 28GHz}
\vspace{-4mm}
\end{figure*}

At the follower layer (Algorithm~\ref{Algorithm_2}), Fig.~\ref{fig:alg2_BR iterations} demonstrates that the per-slot Best Response (BR) iterations required to satisfy the GNE tolerance decrease over time, consistent with best response dynamics and equilibrium power $P_u$ in HN joint power control, and shared constraints. Moreover, the mean GNE gap decays rapidly with BR iteration, confirming contraction of the equilibrium gap within each slot, including the $3$~THz case exhibits a higher initial gap and slower decay, reflecting stronger coupling and a more challenging equilibrium landscape at higher carrier frequencies (Fig. \ref{fig:alg2_gnegap}. 
Finally, at the Bayesian-cooperative refinement layer (Algorithm~\ref{algorithm_3}), Fig.~\ref{fig:alg3_coalition} shows that coalition scaling converges to a steady band near unity within posterior driven entropy $H(p)$ stabilization of sensing-driven localization, following from fig. \ref{fig:powerfraction_entropy}. The cooperative power-update residual norm decays toward zero (log scale) and the total jammer power stabilizes, evidencing convergence of cooperative refinements without destabilizing power escalation depicted in fig. \ref{fig:alg3_power}. Overall, fig. \ref{fig:alg_conv} confirm that iBEAMS achieves hierarchical convergence: the leader stabilizes power split and incentives while tracking secrecy, the follower game reaches GNE efficiently with decreasing iteration burden, and the Bayesian-cooperative layer converges in uncertainty reduction and coalition/power coordination, with the THz regime requiring comparatively stronger and longer adaptation.

Figure~\ref{fig:ibeams at 28GHz} jointly demonstrates that iBEAMS converges to a stable secure operating regime in which the leader’s power-split and incentive prices settle to steady values while the distributed HNs perform sparse, event-triggered THN/JHN role switching that preserves high long-term utility and secrecy.  
Firstly,Fig.~\ref{fig:powerfraction_entropy} illustrates the
leader-layer (Bayesian-Stackelberg) convergence by jointly tracking: (i) sensing uncertainty through the posterior entropy $H(p)$ (left axis), and (ii) the leader's control variables as the BS power-split coefficients $(\alpha_t,\beta_t,\gamma_t)$ and the incentive prices$(\pi_t,\tau_t,\kappa_t)$ (right axis). Over time, the data power fraction $\alpha_t$ monotonically increases and then saturates, indicating that the leader progressively reallocates ISAC BS resources toward information-bearing transmission once a stable secrecy regime is established. In contrast, the artificial-noise share $\beta_t$ gradually decreases, while the sensing share $\gamma_t$ drops more sharply before settling at a small nonzero value, reflecting a transition from an exploration phase to an exploitation regime in which sensing is kept lightweight but persistent. The incentive prices exhibit clear saturation
behavior consistent with convergence: the jamming reward $\pi_t$ rapidly ramps up to its upper bound, signaling that iBEAMS persistently incentivizes cooperative jamming under strong eavesdropping; the leakage penalty $\tau_t$ decays toward zero as leakage is driven into an acceptable range; and the
sensing/information reward $\kappa_t$ increases steadily, promoting continued Bayesian information acquisition even as the direct sensing power fraction $\gamma_t$ is reduced. Meanwhile, the entropy $H(p)$ remains high but bounded,
fluctuating within a narrow band. Taken together, the joint flattening of $(\alpha_t,\beta_t,\gamma_t)$,and $(\pi_t,\tau_t,\kappa_t)$, along with bounded entropy dynamics, demonstrates that the iBEAMS leader layer attains a stable closed-loop operating equilibrium that consistently balances secrecy-driven jamming incentives, leakage mitigation, and sustained yet low-overhead sensing support.
Secondly, Fig.~\ref{fig:roleswitch_bar} characterizes the local-layer adaptation of iBEAMS through time-varying role assignment between THNs and JHNs. The top panel shows that the network operates predominantly in a jammer-dominant regime: most HNs are persistently selected as JHNs (red), while only a small subset remain THNs (green) in each slot. This reflects operation under strong (or worst-case) eavesdropping conditions, where the controller prioritizes secrecy by allocating the majority of edge resources to cooperative jamming, while maintaining a minimal set of THNs to sustain data delivery. Fig. \ref{fig:roleswitch_intensity}  highlights discrete role-switching events (T$\to$J and J$\to$T) that occur
sporadically in response to changes in the environment (active Eve location, secrecy-margin degradation, or proximity-based jammer selection), showing that role updates are event-driven rather than continuous. Furthermore, fig. \ref{fig:roleswitch_utility} demonstrates that these switching events induce only transient perturbations in aggregate HN utility, indicating that switching
instants are followed by rapid recovery to a high, stable state. In parallel, the JHN fraction remains tightly constrained within a narrow band across slots, indicating the convergence of the network-wide security posture to a steady operating point, with role adjustments acting as sparse corrective actions rather than a source of instability.
\begin{figure*}[t]
\centering
\subfloat[]{\includegraphics[width=0.35\textwidth]{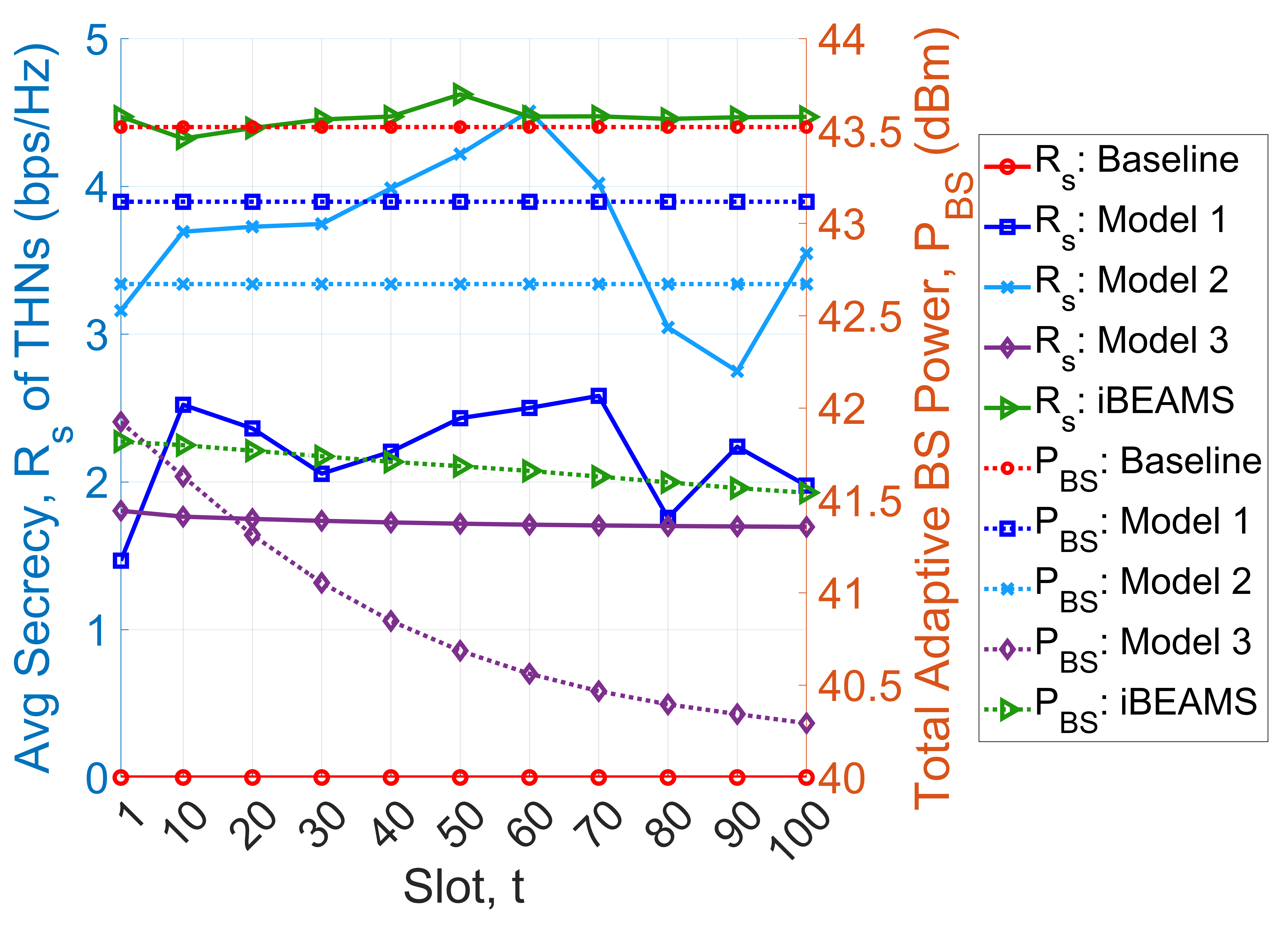}%
\label{fig:power_secrecy_baseline}}
\hfill
\subfloat[]{\includegraphics[width=0.28\textwidth]{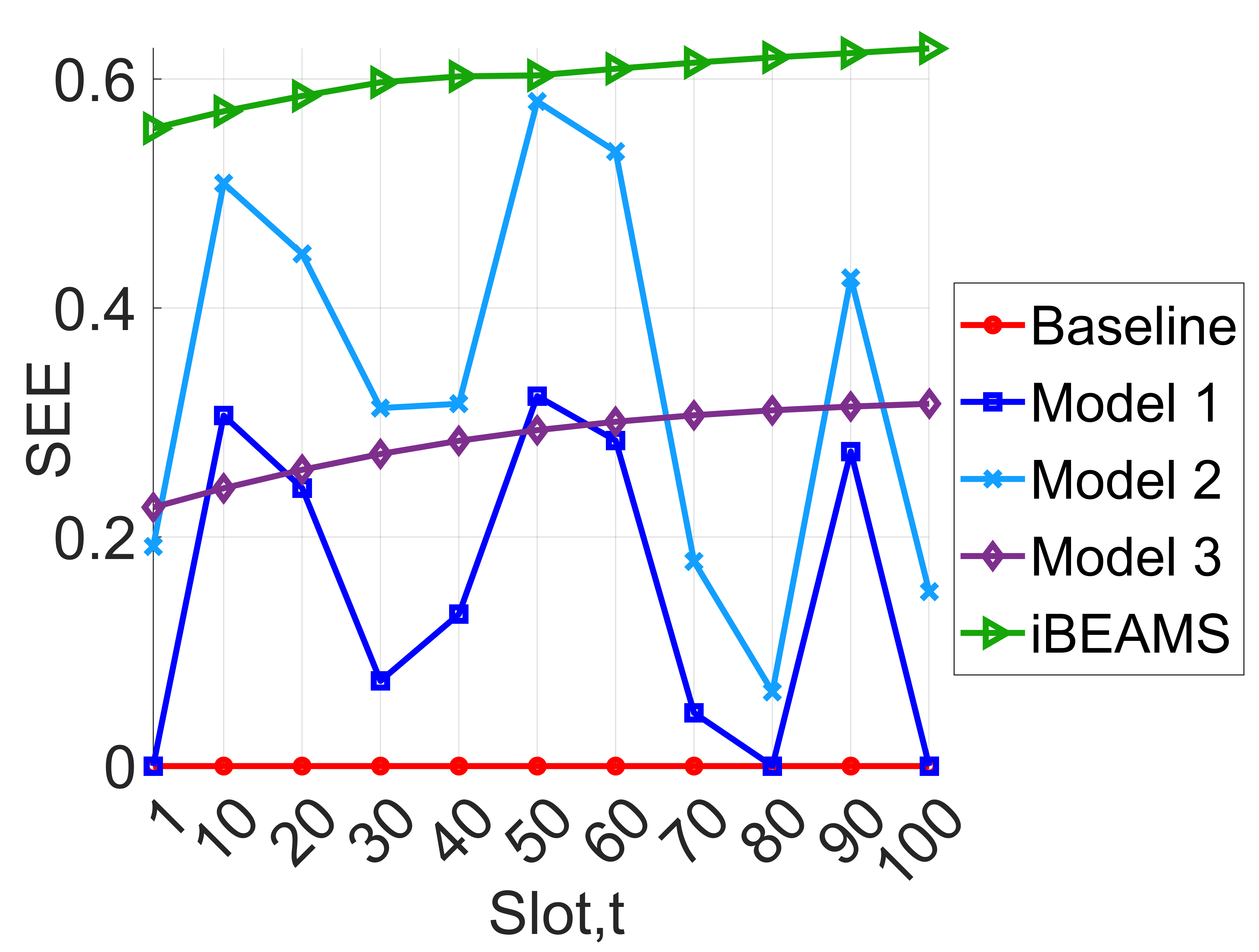}%
\label{fig:SEE_baseline}}
\hfill
\subfloat[]{\includegraphics[width=0.33\textwidth]{Figures/Outagesuccessrate.png}%
\label{fig:outagesuccess_baseline}}
\vspace{-2mm}
\caption{Comparative performance of the proposed iBEAMS unified framework versus baseline and multiple models: (a) power-secrecy trade-offs, (b) secrecy energy efficiency (SEE), and (c) outage probability and secrecy success rate within time slot, $t$ at $28$ GHz}
\label{fig:compare_baseline}
\vspace{-4mm}
\end{figure*}

In Figure~\ref{fig:compare_baseline}, we examine a significant numerical comparison of the proposed iBEAMS architecture against the baseline and three benchmark models. In Fig.~\ref{fig:power_secrecy_baseline}, the baseline exhibits a zero average secrecy rate over all slots, as the system could not communicate data to legitimate users over \emph{the worst-case and strongest eavesdropper}, despite operating the BS around $43.5$~dBm, confirming that conventional communications without cooperative jamming are unable to counter strong eavesdropping. Model~1 with conventional communications and injected fixed AN improves the average secrecy rate to roughly $2$–$2.5$~b/s/Hz, while Model~2 (ISAC BS with Stackelberg decision) further increases it to about $3.5$–$4.0$~b/s/Hz. However, both require comparable or higher BS power levels (around $42$–$43$~dBm). Model~3 (ISAC BS with Stackelberg decision and HN with role switching) reduces the BS power from approximately $41.8$~dBm to near $40.2$~dBm over time, but at the cost of a relatively modest secrecy rate of $\approx 1.6$–$1.8$~b/s/Hz. In contrast, iBEAMS stabilizes the average secrecy rate of THNs around $4.4$–$4.7$~b/s/Hz while maintaining a nearly flat BS power trajectory close to $41.5$~dBm. Hence, relative to the best benchmarks, iBEAMS delivers a roughly $15$–$25\%$ higher secrecy rate with lower transmission power, demonstrating a strictly superior secrecy–power trade-off.

The corresponding secrecy energy efficiency (SEE) profiles in Fig.~\ref{fig:SEE_baseline} reinforce this trend. The baseline again attains an SEE of zero bits/Joule, whereas Model~1 achieves only $\approx 0.15$–$0.25$~bits/J and Model~3 hovers around $0.25$–$0.32$~bits/J. Although Model~2 is more efficient, it fluctuates mostly in the $0.35$–$0.55$~bits/J range. By contrast, iBEAMS consistently operates near the upper envelope of all curves, sustaining SEE values between $0.55$ and $0.63$~bits/J for the majority of slots, with only occasional transient drops associated with role-switching events. Overall, iBEAMS offers roughly a 2$\times$ SEE gain over Model~1 and a clear margin on the order of $30$–$70\%$ over Model~2, while simultaneously guaranteeing higher secrecy rates.
Finally, Fig.~\ref{fig:outagesuccess_baseline} summarizes the reliability of secrecy in terms of outage probability and secrecy success rate. Consequently, the baseline experiences a $100\%$ secrecy outage and thus $0\%$ success rate at the chosen threshold, indicating that it never meets the required secrecy level. Model~1 reduces the outage probability to $32.5\%$, and $67.5\%$ success, and Model~2 further improves to $12.5\%$ outage ($87.5\%$ success). Both Model~3 and iBEAMS achieve zero outage and $100\%$ secrecy success across the simulated horizon; however, when combined with the higher secrecy rates and substantially better SEE shown in Figs.~\ref{fig:power_secrecy_baseline}, \ref{fig:SEE_baseline}, iBEAMS clearly dominates Model~3. Therefore, the three metrics jointly demonstrate that iBEAMS attains outage-free, high-rate secrecy with markedly superior energy efficiency, thereby establishing a strong performance advantage over all baseline and benchmark designs.
\begin{figure*}[t]
\centering
\subfloat[]{\includegraphics[width=0.35\textwidth]{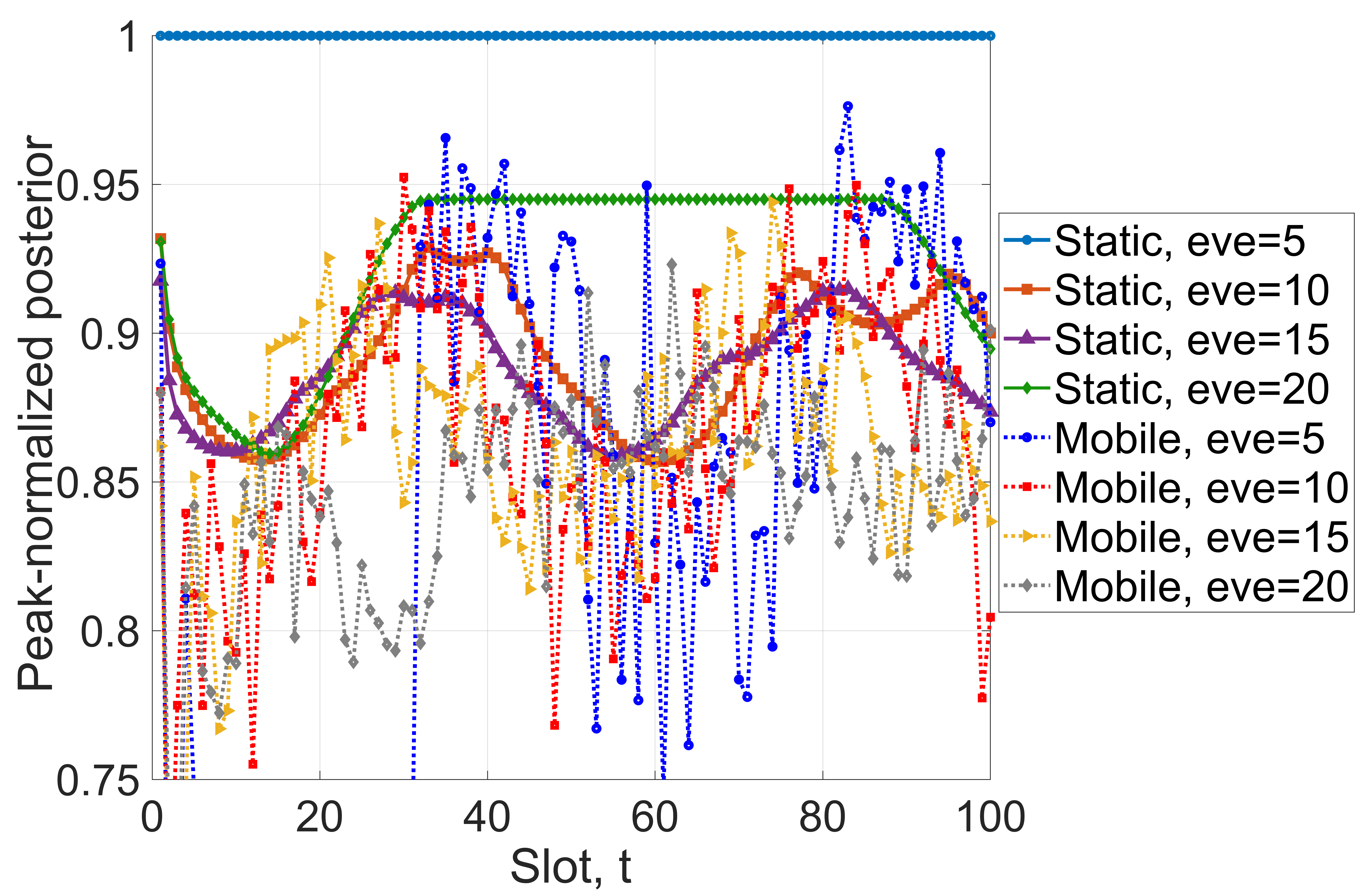}%
\label{overall_plots_eve}}
\hfill
\subfloat[]{\includegraphics[width=0.30\textwidth]{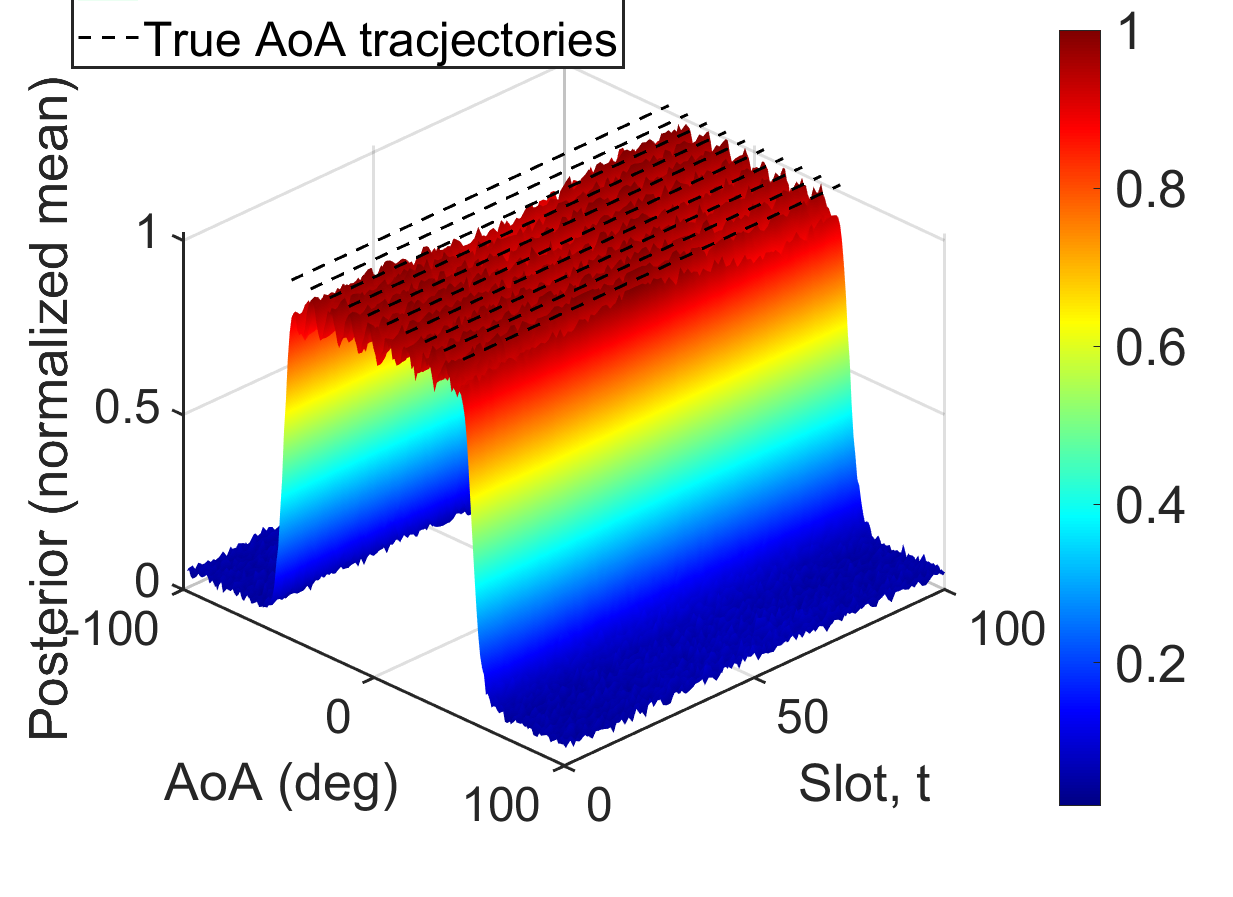}%
\label{fig:posterior heatmap static eve}}
\hfill
\subfloat[]{\includegraphics[width=0.30\textwidth]{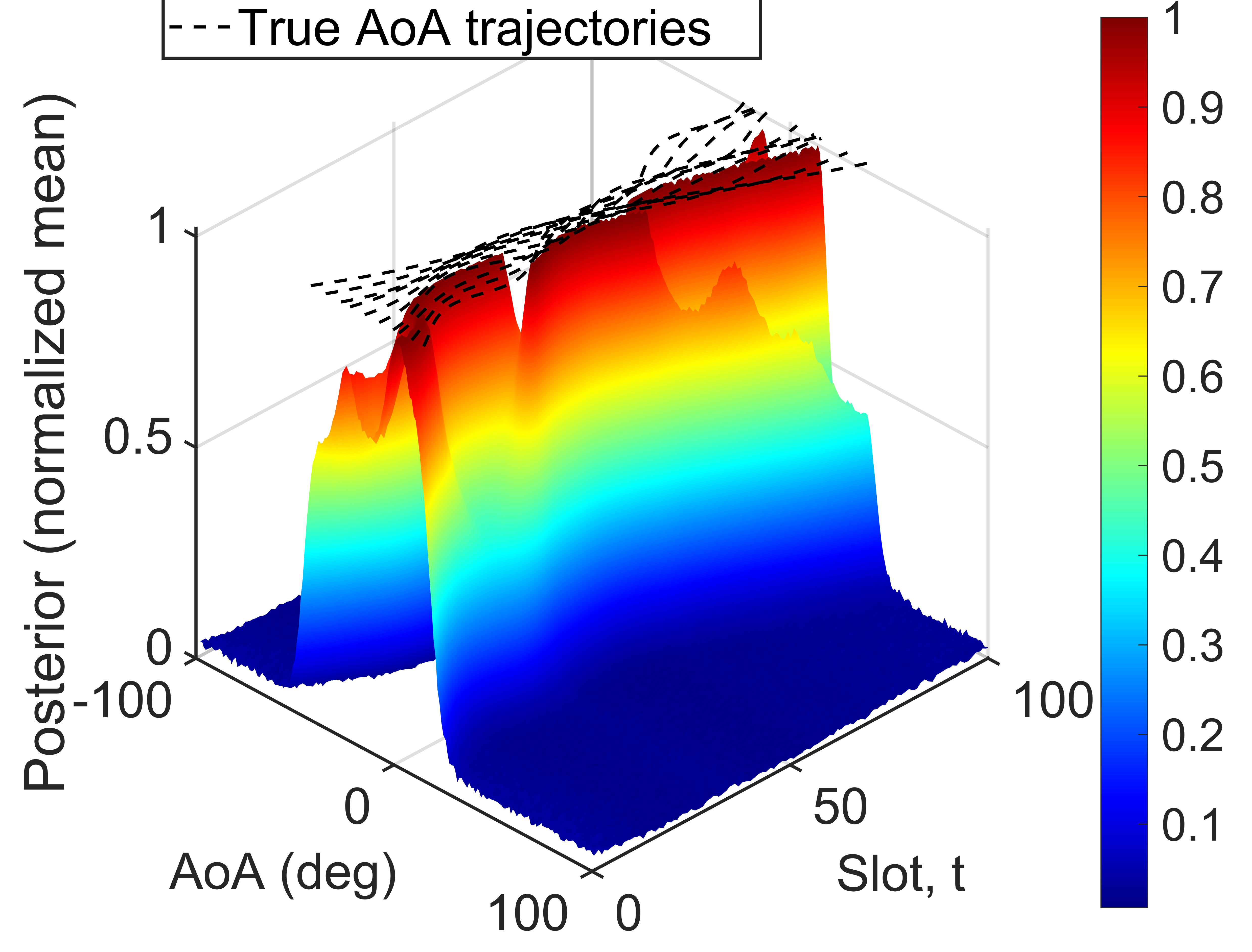}%
\label{fig:posterior heatmap mobile eve}}
\vspace{-2mm}
\caption{Comparison of Eve AoA posterior heatmaps for the proposed iBEAMS framework under (a) quasi-static and (b) mobile eavesdropping scenarios at $28$~GHz, where each panel reports the time-varying posterior $p(\theta \mid t)$ for four representative eavesdroppers}
\label{fig:posterior_heatmap}
\vspace{-4mm}
\end{figure*}

Figures~\ref{fig:posterior_heatmap} illustrate the dynamics of the AoA posterior $p(\theta\!\mid\!t)$ for four representative eavesdroppers (Eve~5, Eve~10, Eve~15, and Eve~20) under a static and mobile deployment.  For all cases, the iBEAMS Bayesian layer rapidly concentrates the posterior around the true AoA and then maintains a narrow, high–confidence ridge over time (fig. \ref{overall_plots_eve}).  After an initial learning phase of roughly $t \approx 10$–$15$ slots, the dominant
mode of $p(\theta\!\mid\!t)$ stabilizes within a tight angular window of approximately $\pm 5^\circ$–$\pm 10^\circ$ around the true direction. This behavior is consistent across all four network configurations (from 5 up to 20 Eves), demonstrating that the proposed sensing-and-learning mechanism can accurately infer and \emph{retain} the correct posterior even as the number of potential eavesdroppers increases.In addition,  fig.~\ref{fig:posterior heatmap static eve} demonstrates the true AoA posterior heatmap $p(\theta\!\mid\!t)$ for the same set of representative Eves when the eavesdroppers are static and the number of Eves is around~$10$. 
Similarly, fig. ~\ref{fig:posterior heatmap mobile eve}, the heatmaps exhibit a sequence of distinct angular tracks rather than a single stationary ridge, reflecting changes in the dominant AoA segment as the active Eve moves or as different Eves become most threatening.  Importantly, at each segment the posterior still
forms a sharply peaked lobe, with peak probabilities again reaching $\approx 0.8$–$0.9$, and the transition between segments occurs within a small number of slots.  
This indicates that iBEAMS is able to quickly diminish outdated directions and re–localize the active eavesdropper whenever the channel geometry changes.Likewise, this behavior is preserved across scenarios with 5, 10, 15, and 20 candidate Eves highlights the robustness of the hierarchical Bayesian update: even when the pool of adversaries grows, the framework continues to deliver high–confidence, angle–selective posteriors that
track the true direction over time.  Overall, figure~\ref{fig:posterior_heatmap} comparison between the static and mobile heatmaps confirms that iBEAMS not only learns an accurate static posterior but also maintains reliable AoA tracking in highly dynamic environments, thereby enabling superior, geometry-aware physical-layer security against both few and many eavesdroppers.

\begin{figure}[htp]
\centering
\includegraphics[width=0.48\textwidth]{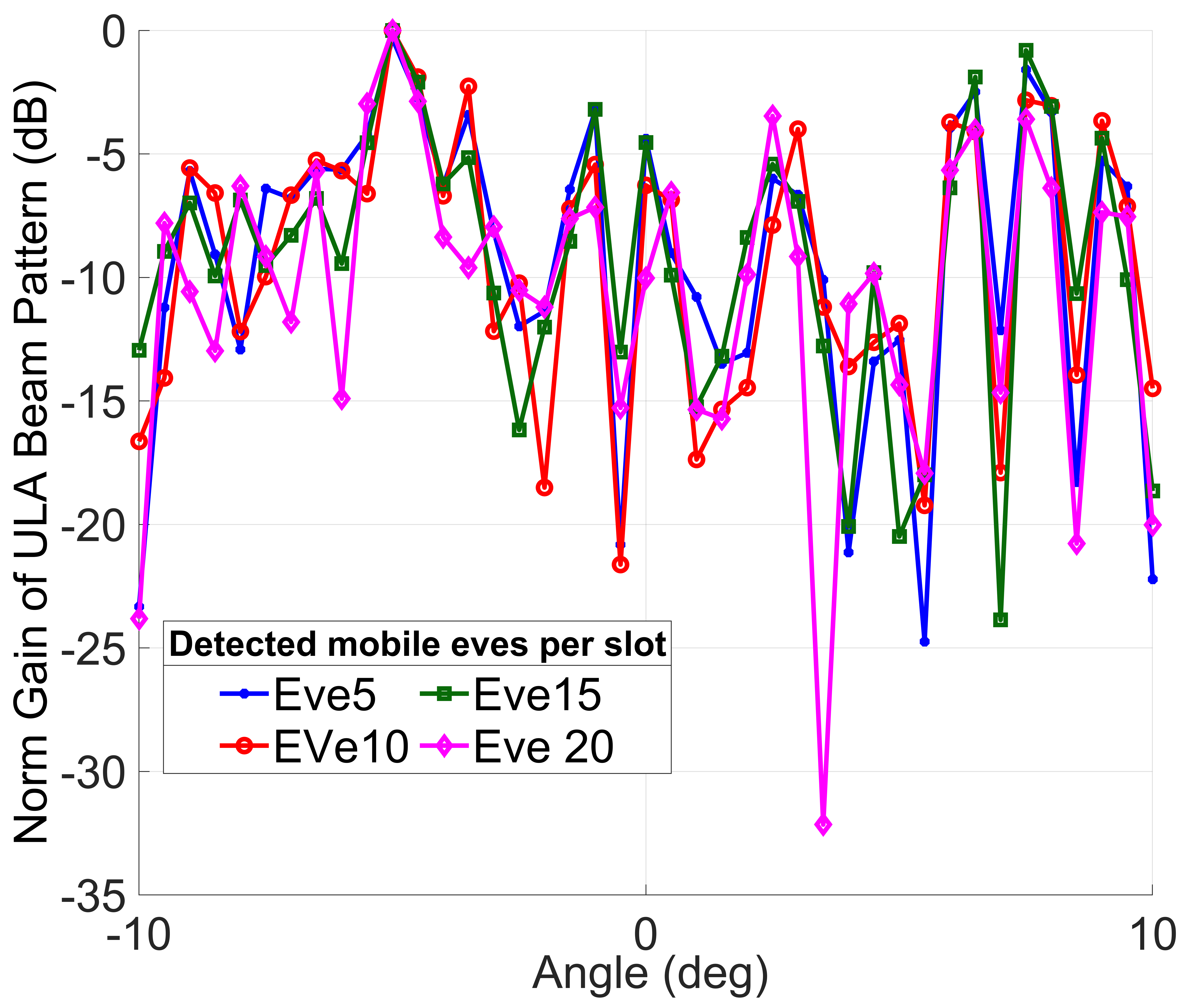}
\caption{Normalized ULA beam pattern at $28$~GHz for posterior–aware iBEAMS beamforming, showing the array gain versus angle for four representative mobile eavesdroppers within the $\pm 10^\circ$ sector}
\label{ULA_beampattern over few to high eves}
\end{figure}
Figure~\ref{ULA_beampattern over few to high eves} illustrates the normalized ULA jamming beam pattern synthesized by the proposed iBEAMS framework when steering toward four representative mobile eavesdroppers (Eve5, Eve10, Eve15, and Eve20) whose AoA posteriors are provided by the Bayesian leader layer. Across the angular window $\theta\in[-10^\circ,10^\circ]$, all four traces exhibit a sharp main lobe with peak gain close to $0$~dB around the posterior mean AoA, while neighboring directions are consistently attenuated by about $5$–$15$~dB. In several angles the array forms deep nulls below $-25$ to $-30$~dB, demonstrating that the CSI–aware, posterior-driven precoder can shape pronounced spatial notches in unintended directions while concentrating power precisely along the estimated eavesdropper direction. The strong overlap between the four curves indicates that, even as the active Eve index changes and the number of eavesdroppers increases or decreases, iBEAMS rapidly re-steers the jamming beam to the new posterior without sacrificing array gain or side-lobe suppression. Consequently, the beam pattern faithfully reflects the underlying near-field physical channel under CSI bounds and confirms that iBEAMS can maintain highly directive, mobile-Eve–aligned jamming beams, thereby strengthening physical-layer protection with minimal leakage toward benign directions.
\section{Conclusions}
\label{Conc.}
To conclude, this research has introduced iBEAMS, a multi-layer hierarchical Stackelberg–GNE–Bayesian framework for secure and energy-efficient ISAC in networks populated by hybrid edge nodes. The architecture tightly couples a Bayesian–Stackelberg ISAC BS controller, a GNE-based hybrid-edge node's power and role-selection game, and a Bayesian cooperative refinement layer that shapes geometry-aware jamming. Numerical evaluations from $28$~GHz up to $3$~THz demonstrate hierarchical convergence: the leader’s power-splitting variables and incentive prices settle around stable operating points with small secrecy-tracking error, the follower game reliably attains a GNE with a diminishing equilibrium gap, and the cooperative layer drives the AoA posterior entropy, coalition scaling, and aggregate jammer power toward steady regimes without destabilizing oscillations. Relative to conventional communication, fixed-AN, and Stackelberg-only or role-switching benchmarks, iBEAMS achieves substantially higher secrecy rates (approximately $4.4$–$4.7$~b/s/Hz at $28$~GHz), about a $2\times$ improvement in SEE over fixed AN, and a $30$–$70\%$ SEE gain over Stackelberg-only ISAC, while maintaining zero secrecy outage. The Bayesian layer further yields narrow, high-confidence AoA posteriors and sharply directive, posterior-aligned jamming beams in both static and mobile eavesdropping scenarios, enabling robust, geometry-aware physical-layer protection.

Notwithstanding these benefits, several limitations remain. The current formulation is restricted to a single-cell ISAC BS with a moderate number of hybrid nodes and does not explicitly capture inter-cell coordination, multi-BS coupling, or ultra-dense deployments. Again, RF and wideband propagation effects are modeled at a simulation and theoretical level; a more faithful treatment of hardware impairments, beam alignment, phase-noise, and nonlinear front-end distortions is required for hardware-accurate design. Moreover, the algorithms further needs enhancement of a rigorous stability and robustness analysis under model mismatch, rapid mobility, and adversarial behavior. Finally, although computational and signaling overheads are quantified numerically, real-time validation on mmWave/THz ISAC testbeds and the development of lower-complexity of iBEAMS remain essential steps toward deployment. Taken together, the results indicate that iBEAMS offers a principled pathway for addressing the core challenge of securing mobile eavesdroppers in networks of distributed hybrid edge nodes under stringent power and QoS constraints.
\vspace{-0.2cm}
\appendices
\section{Follower Utility and GNE Formulation}
\label{app:follower_utility_gne}

In slot $t$, each active HN $u\in\mathcal{A}(t)$ selects a transmit power $P_u(t)\in[0,P_u^{\max}]$ and a role $\mathrm{role}_u(t)\in\{\mathrm{THN},\mathrm{JHN}\}$ in response to the leader's transmission
$\mathbf{a}_t=(\alpha_t,\beta_t,\gamma_t,\pi_t,\tau_t,\kappa_t)$.
To align individual incentives with the SEE objective, the local utility of HN $U$ is defined as
\begin{equation}
\begin{aligned}
U_u\!\big(P_u(t),\mathrm{role}_u(t);\mathbf{a}_t,\mathbf{P}_{-u}(t)\big)
&= \eta_u R_s(u,t) - c_u P_u(t) \\
&\quad - \tau_t \Xi_u(t) + \pi_t J_u(t) + \kappa_t I(t),
\end{aligned}
\label{eq:HN_utility_app}
\end{equation}
where $\mathbf{P}_{-u}(t)$ collects the powers of all other HNs,  $\eta_u R_s(u,t)$ rewards secrecy throughput (effective when HN acts as a THN), $c_uP_u(t)$ penalizes power expenditure, $\tau_t\Xi_u(t)$ penalizes secrecy leakage attributable to HN, $\pi_t J_u(t)$ rewards effective friendly-jamming contributions when HN acts as a JHN, and $I(t)$ denotes the entropy reduction or information gain from ISAC sensing shared among HNs, scaled by the information price $\kappa_t$. Thus, $(\pi_t,\tau_t,\kappa_t)$ implements the proposed SEE aware pricing by embedding BS priorities into follower payoffs.
In addition, HN decision making and utility are constrained by local bounds and network-wide coupling conditions:
\begin{align}
0 \le P_u(t) &\le P_u^{\max}, \quad \forall u, \\
\sum_{u\in\mathcal{A}(t)} P_u(t) &\le P_{\mathrm{FJ}}^{\max}, \\
\Xi_u(t) &\le \Xi_{\max}, \quad \forall u, \\
g\big(\mathbf{P}(t)\big) &\le 0
\end{align}

which defines the global feasible set and the individual feasible set
$\mathcal{P}_u(t)=\{P_u(t)\mid \exists\,\mathbf{P}_{-u}(t)\ \text{s.t.}\ (P_u(t),\mathbf{P}_{-u}(t))\in\mathcal{P}(t)\}$.
Because $U_u$ depends on $\mathbf{P}_{-u}(t)$ and $\mathcal{P}_u(t)$ is formed by shared constraints, the follower interaction forms a GNE game. Thus, a power profile $\mathbf{P}^\star(t)$ is a GNE if, for every $u$ that can be expressed as following the problem from \eqref{eq:prbfollowerutility},
\begin{equation}
U_u\!\big(P_u^\star(t),\mathbf{P}_{-u}^\star(t)\big)\ \ge\
U_u\!\big(P_u(t),\mathbf{P}_{-u}^\star(t)\big),\quad \forall\,P_u(t)\in\mathcal{P}_u(t),
\label{eq:GNE_app}
\end{equation}
where, algorithm~\ref{Algorithm_2} computes an implementable approximation through best-response iterations on a discrete grid:
\begin{equation}
P_u^{(i)}(t)=\arg\max_{P_u(t)\in\mathcal{P}_u(t)}
U_u\!\big(P_u(t),\mathbf{P}_{-u}^{(i-1)}(t)\big),
\label{eq:BR_app}
\end{equation}
terminated when $\|P_u^{(i)}(t)-P_u^{(i-1)}(t)\|\le\epsilon$ for all $u$, yielding $\mathbf{P}^\star(t)$ used for subsequent role switching in Layer~2.
\section{Unified Three-Layer Utility}
\label{app:unified_three_layer_utility}
To formalize the total utility of the three-layer iBEAMS loop in a single objective form, we define the slot-$t$ decision tuple
\(\mathbf{x}_t \triangleq \big(\mathbf{a}_t,\mathbf{z}_t,\mathbf{b}_t\big)\),where (i) the leader variables are
\(\mathbf{a}_t=(P_{\mathrm{BS}}(t),\alpha_t,\beta_t,\gamma_t,\pi_t,\tau_t,\kappa_t)
\),(ii) the \emph{follower} variables are
\(\mathbf{z}_t=(\mathbf{P}(t),\boldsymbol{\rho}(t))
\)with $\mathbf{P}(t)=\{P_u(t)\}_{u\in\mathcal{A}(t)}$ and roles $\boldsymbol{\rho}(t)=\{\mathrm{role}_u(t)\}_{u\in\mathcal{A}(t)}$,
and (iii) the Bayesian belief state is
\(\mathbf{b}_t \triangleq p_t(\theta_E)\)

\begin{equation}
\begin{aligned}
\mathcal{J}_t(\mathbf{x}_t)
\triangleq\;
&\underbrace{\mathrm{SEE}(t)}_{\text{secrecy--energy objective}}
-\lambda_{\mathrm{sec}}\!\left(R_s^{\star}-\overline{R}_s(t)\right)_+ \\
&-\lambda_H\!\left(H(\mathbf{b}_t)-H^{\star}\right)_+
+\lambda_I\,\underbrace{I(t)}_{\text{information gain}},
\end{aligned}
\label{eq:unified_slot_utility}
\end{equation}
where $\overline{R}_s(t)$ is the average secrecy rate of the system, $H^{\star}$ is the target entropy level. For the three-layer coupling and equilibrium response, given $\mathbf{a}_t$ and $\mathbf{b}_t$, the follower layer computes a GNE response from \eqref{eq:HN_utility_app}, and $\mathbf{z}_t^{\star}(\mathbf{a}_t,\mathbf{b}_t)\in \mathrm{GNE}\big(\mathbf{a}_t,\mathbf{b}_t\big),$ under the local and coupling feasibility constraints. The Bayesian layer updates the belief state using an ISAC sensing/refinement operator$\mathbf{b}_{t+1}=\mathsf{B}\!\left(\mathbf{b}_t,\mathbf{a}_t,\mathbf{z}_t^{\star}\right),$ that induces$H(\mathbf{b}_{t+1})$,and $I(t)$.
The overall three-layer control problem can therefore be expressed as the leader optimization,
\begin{equation}
\max_{\mathbf{a}_t\in\mathcal{A}_{\mathrm{lead}}}\;
\mathcal{J}_t\!\Big(\mathbf{a}_t,\mathbf{z}_t^{\star}(\mathbf{a}_t,\mathbf{b}_t),\mathbf{b}_t\Big),
\label{eq:unified_stackelberg_objective}
\end{equation}
where $\mathcal{A}_{\mathrm{lead}}$ collects the feasibility constraints of the leader (power budget and simplex of power-splitting for $(\alpha_t,\beta_t,\gamma_t)$, along with the limits in $(\pi_t,\tau_t,\kappa_t)$). Equation~\eqref{eq:unified_stackelberg_objective} provides a single objective view of iBEAMS: the leader chooses $\mathbf{a}_t$ to maximize SEE while penalizing the secrecy deficit and excessive belief entropy. 
\vspace{-0.2cm}
\section{Cooperative Jamming Optimization in Layer 3}
\label{app:coop_jam_derivation}
In Layer~3, the posterior update $p_t(\theta)$ identifies the most likely eavesdropper directions and enables cooperative refinement among jammer HNs. After the Layer~2 GNE, the THN/JHN roles and a feasible power profile $\mathbf{P}^\star(t)$ are available. Let $\mathcal{C}\subseteq\mathcal{J}(t)$ denote a jamming coalition (a subset of JHNs) selected based on the dominant posterior peaks. In Layer~3, only the coalition powers $\mathbf{p}_{\mathcal{C}}(t)\triangleq\{P_k(t)\}_{k\in\mathcal{C}}$ are refined, while all other powers remain fixed at their current values.
\(\mathbf{p}_{-\mathcal{C}}(t)=\mathbf{p}^\star_{-\mathcal{C}}(t)\)

The refinement criterion is to improve the aggregate secrecy of the currently served THNs under fixed leader parameters and fixed non-coalition actions. Using the standard definition of $R_s$, the coalition-level refinement is obtained by restricting the optimization variables to $\mathbf{p}_{\mathcal{C}}(t)$ while keeping all other decisions fixed. This yields the Layer~3 cooperative-jamming subproblem:
\begin{equation}
\mathbf{p}_{\mathcal{C}}^{\mathrm{new}}(t)
=\arg\max_{\mathbf{p}_{\mathcal{C}}(t)}\;
\sum_{u\in\mathcal{T}(t)} R_s\!\big(u,t;\mathbf{p}_{\mathcal{C}}(t),\mathbf{p}_{-\mathcal{C}}(t)\big),
\label{eq:app_coalition_obj}
\end{equation}
where $\mathcal{T}(t)$ denotes the THN set, and demonstrates that Layer~3 performs a local improvement step in the objective of network secrecy conditioned on the equilibrium outcome of Layer~2 and the update of belief of Layer~3.
Since friendly jamming may harm legitimate receivers through residual leakage, refinement is performed under THN-protection constraints. A common leakage model is an interference-temperature bound at each THN $u$ due to the coalition:
\begin{equation}
\Xi_u(t)\;\triangleq\;\sum_{k\in\mathcal{C}} P_k(t)\,|h_{k\rightarrow u}(t)|^2 \;\le\; \Xi_{\max},\qquad \forall u\in\mathcal{T}(t),
\label{eq:app_leak_constraint}
\end{equation}
where $h_{k\rightarrow u}(t)$ is the channel from jammer $k$ to THN $u$. This constraint enforces QoS protection while allowing coalition power redistribution.

To ensure that the refined jamming is aligned with the updated belief $p_t(\theta)$, a posterior-weighted shaping constraint is imposed on the coalition radiation towards the angular grid $\Phi$. Let $g_k(\theta)\ge 0$ denote the directional gain of the jammer $k$ (beampattern) toward $\theta$. The coalition jamming field is
$J_{\mathcal{C}}(\theta,t)=\sum_{k\in\mathcal{C}} P_k(t)\,g_k(\theta)$
Moreover, a belief-aligned shaping requirement can be expressed as a lower bound on posterior-weighted jamming energy,
$\sum_{\theta\in\Phi} p_t(\theta)\,J_{\mathcal{C}}(\theta,t)\;\ge\; J_{\min}(t),$ or, equivalently, by enforcing sufficient jamming on the dominant posterior support ( such as $\theta\in\Phi_{\mathrm{peak}}(t)$ where $p_t(\theta)$ is large). This ensures that refinement concentrates energy in secrecy-critical directions identified by sensing.
Combining the coalition objective with the local power limits, THN leakage protection, and posterior-weighted shaping yields the Layer~3 problem used in Algorithm~\ref{algorithm_3}:
\begin{equation}
\scriptsize
\begin{aligned}
\{P_k^{\mathrm{new}}(t)\}_{k\in\mathcal{C}}
&=\arg\max_{\{0\le P_k(t)\le P_k^{\max}\}_{k\in\mathcal{C}}}
\sum_{u\in\mathcal{T}(t)} 
R_s\!\big(u,t;\mathbf{p}_{\mathcal{C}}(t),\mathbf{p}_{-\mathcal{C}}(t)\big)\\
\text{s.t.}\quad
&\sum_{k\in\mathcal{C}} P_k(t)\,|h_{k\rightarrow u}(t)|^2 
\le \Xi_{\max},\ \forall u\in\mathcal{T}(t),\\
&\sum_{\theta\in\Phi} p_t(\theta)
\sum_{k\in\mathcal{C}} P_k(t)\,g_k(\theta)
\ge J_{\min}(t).
\end{aligned}
\label{eq:app_coalition_opt_full}
\end{equation}
Problem~\eqref{eq:app_coalition_opt_full} is generally nonconvex. Hence, it is implemented in Layer~3 through low-complexity refinement, including discrete-grid search, successive convex approximation, or projected best-response updates, and it is terminated when the secrecy-improvement criterion in Algorithm~\ref{algorithm_3} is satisfied.
\bibliographystyle{IEEEtran}
\bibliography{reference}
\end{document}